\documentclass[aps,12pt]{revtex4-2}
\usepackage{amsmath}
\usepackage{amssymb}
\usepackage{bm}
\usepackage{graphicx}
\usepackage{epsf}
\usepackage{xcolor}

\usepackage{color}

\begin{document}

\title{Ballistic transport of a polariton ring condensate with spin precession}

\author{Q. Yao$^1$, E. Sedov$^{2,3,4,5}$, S. Mukherjee,$^6$ J. Beaumariage$^1$, B. Ozden,$^7$ K. West,$^8$ L. Pfeiffer,$^8$\\
A. Kavokin,$^{2,3,4,5,9}$ and  D. W. Snoke$^1$}

\affiliation{$^1$Department of Physics and Astronomy, University of Pittsburgh\\ 3941 O'Hara St., Pittsburgh, PA 15260, USA\\
$^2$Key Laboratory for Quantum Materials of Zhejiang Province, School of Science, Westlake University, 18 Shilongshan Rd, Hangzhou 310024, Zhejiang, China\\
$^3$Westlake Institute for Advanced Study, 18 Shilongshan Rd, Hangzhou 310024, Zhejiang, China\\
$^4$Spin Optics Laboratory, St. Petersburg State University, Ulyanovskaya 1, St. Petersburg 198504, Russia\\
$^5$Russian Quantum Center, 100 Novaya St., Skolkovo, Moscow region, 143025, Russia\\
$^6$Joint Quantum Institute, University of Maryland and National Institute of Standards and Technology, College Park,
Maryland 20742, USA\\
$^7$Department of Physics, Penn State Abington, Abington, PA 19001, USA\\
$^8$Department of Electrical Engineering, Princeton University, Princeton, NJ 08544, USA \\
$^9$Moscow Institute of Physics and Technology, Dolgoprudnyi, Moscow Region, 141701, Russia
}

\begin{abstract} 
It is now routine to make Bose-Einstein condensates of polaritons with long enough lifetime and low enough disorder to travel ballistically for hundreds of microns in quasi-one-dimensional (1D) wires. We present observations of a non-equilibrium polariton condensate injected at one point in a quasi-1D ring, with a well-defined initial velocity and direction. A clear precession of the circular polarization is seen, which arises from an effective spin-orbit coupling term in the Hamiltonian. Our theoretical model accurately predicts the experimentally observed behavior, and shows that ``zitterbewegung'' behavior plays a role in the motion of the polaritons.
\end{abstract}

\maketitle



Ring condensates have been studied in several systems (e.g., Refs. \cite{ring1,ring2,ring3,ring4}), including in exciton-polariton condensates \cite{pnas}, where it was shown that quantum phase coherence around the ring led to quantized circulation analagous to quantized vortices in liquid helium. Several other experiments \cite{lagou,baum,perovskite,polarrings,polarcurrents} have also studied polariton condensates in a ring geometry, but primarily showed spatial pattern formation due to an incoherent mechanism, namely an instability in the interaction between the condensate and a thermal reservoir, leading to a natural length scale for self-trapping \cite{tureci}. 

In the present work, we report direct imaging of a polariton condensate as it moves ballistically and coherently in a circular ring. In these experiments, the initial velocity and direction of the condensate is determined by the position of the point of injection of the condensate into the ring. Although the condensate is in a single energy state, its polarization precesses around the ring, because of a term in the Hamiltonian analogous to the spin-orbit coupling which also gives ``zitterbewegung'' \cite{z1,z2,z3,z4,z5,z6}, that is, side-to-side motion of a ballistic wave packet. Because the polaritons have extremely light mass, of the order of $10^{-4}m_0$, where $m_0$ is the free-space mass of an electron, they have long wavelength at easily obtainable temperatures, which makes the ballistic paths large enough (tens of microns) to see by standard optical imaging.

{\bf  Experimental methods}.  The microcavity sample was grown on a GaAs (001) substrate by molecular beam epitaxy (MBE). It consists of a $3\lambda/2$ microcavity and three sets of four GaAs quantum wells placed at the antinodes of the cavity photon mode. Two AlGaAs/AlAs distributed Bragg reflectors (DBRs) were used to create the microcavity, with 32 periods in the top DBR and 40 periods in the bottom DBR. Individual pieces measuring 5 mm x 5 mm from the center of the wafer were used to construct the ring structures. It is worth noting that the wafer's centers are flatter than the remainder, allowing for consistent photon energy across the ring structures. The detuning of the ring is $-12.38$~meV, corresponding to 76\% photonic fraction. The Rabi splitting is $\sim19.6$~meV according to the characterization results before the sample was etched. The array of ring structures with a radial width of approximately 15~$\mu\text{m}$ and a diameter of 100~$\mu\text{m}$ was fabricated using standard photolithography techniques. The top DBR was dry-etched to ring shape with a 20:7 $\textrm{BCl}_3/\textrm{Cl}_2$ inductively coupled plasma (ICP) reactive ion etch (RIE) at 3.0~mT chamber pressure, 600 W ICP power, and 75 W RF bias power. 

During the experiment, a continuous-wave Ti:Sapphire laser, actively locked at $E_{pump} = 1705.6$~meV ($\lambda = 727$~nm), was used to non-resonantly pump the sample  through a 20$\times$0.4-N.A. microscope objective. A mechanical chopper with a 1.7\% duty cycle at 400 Hz was placed in the laser path to reduce heating. All reported powers are the peak power of each short pulse. The pump spot was $\approx 8 \,\mu\text{m}$ FWHM. The same microscope objective was used to capture photoluminescence (PL) from the ring of 1593~meV (778~nm). The reflected laser was blocked by an energy filter (750~nm long-pass filter). Both real-space images and energy-resolved images were taken with a charged-coupled device (CCD) camera. We also obtained angle-resolved imaging by switching the lens combination. Half waveplates, quarter waveplates, and polarizers were utilized for polarization measurement, which is similar to that in Ref.~\cite{shouvik2}. All measurements were performed by cooling the microcavity to low temperature (below 10 K) in a continuous-flow cold-finger cryostat. Details of experimental setup can be found in Supplemental Information \cite{suppl}. The images were normalized by camera integration time, image counts and the power of the neutral density filter in front of the camera. The normalized image looks the same as the raw data with proper color scales.

{\bf Experimental approach and results}. Exciton-polaritons in a microcavity, called here simply polaritons, are mixed states of excitons and photons, which have the property of repulsive interactions due to their excitonic component, while having very light mass due to their photonic component. Numerous experiments have shown the effect of Bose-Einstein condensation of polaritons (e.g., Refs. \onlinecite{d2006,s2007}; for a review see Ref.~\onlinecite{ubec}.)  The polariton condensate is typically created at bath temperatures of 4~K; in our experiment, the condensate is out of equilibrium but still has long-range phase coherence, although under proper conditions of confinement it can reach equilibrium \cite{prl}. 

\begin{figure}
	\centering
	\includegraphics[width=0.9\textwidth]{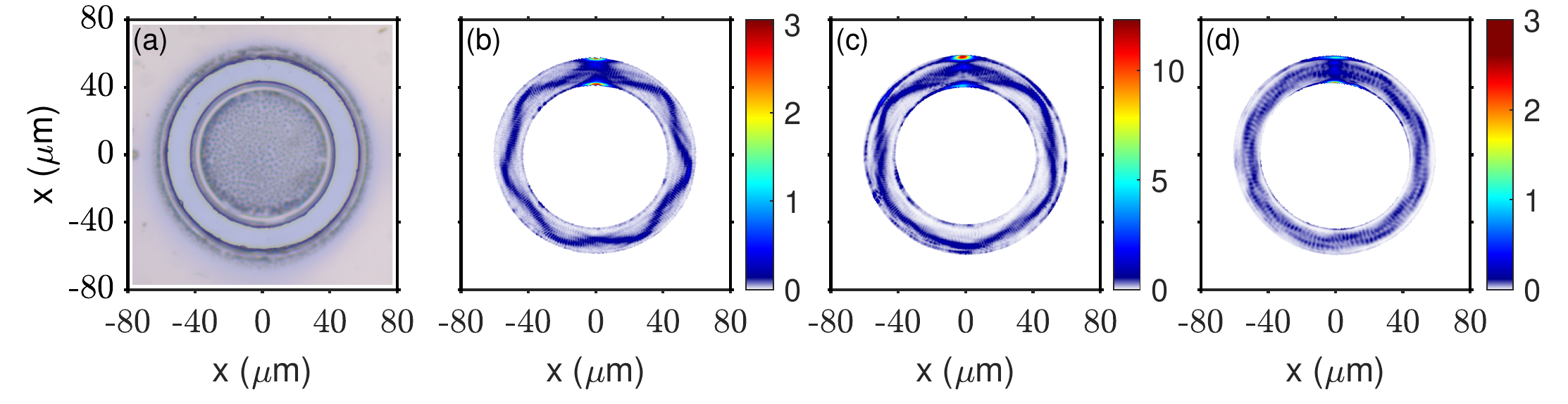}
	\caption{(a) Optical image of typical ring structure. (b)--(d) A series of intensity-normalized real-space images of the light emission from polariton condensate in a ring trap, with slightly different pump positions (the pump is moved in the radial direction). The pump power was about $6\it P_{\rm th}$, where $P_{\rm th}$ was the critical threshold for Bose condensation.}
	\label{fig:pos}
\end{figure}

For these experiments we used an AlGaAs/GaAs microcavity structure with very high $Q$ (ca.~300000), very similar to the structures used in prior experiments \cite{nelsen,steger} which showed ballistic motion of polaritons over hundreds of microns.  These microcavities were etched into ring structures with a radial width of approximately 15~$\mu$m and a diameter of 100~$\mu$m. Figure \ref{fig:pos}(a) shows a typical structure. In prior experiments with polaritons in rings like these \cite{shouvik1,shouvik2}, there was a gradient of potential energy of the polaritons across the rings, which caused the polaritons to eventually settle into the lowest-energy point on one side of the ring. In the present experiments, great care was taken to have a very flat potential-energy profile around the rings, so that the polaritons flow freely around the ring without attraction to any one location. 

Residual strain from the etching process creates a strain profile in the rings in the radial direction that leads to a potential-energy barrier that deters the polaritons from leaking out the sides. 
Therefore the polaritons are confined to a quasi-one-dimensional motion in a narrow channel around the ring. The potential energy profile felt by the polaritons in the radial direction is approximately harmonic, giving a ladder of quantized states for radial motion separated by about 200~$\mu$eV, well below the value of $k_BT$ for the effective temperature of the polaritons, which is about 10~K. (See the Supplementary Information \cite{suppl} for further details about the radial potential-energy profile). The polariton condensate therefore can have motion in the radial direction.

For these experiments we generated the polaritons by focusing a non-resonant pump laser onto a small spot on one side of the ring. Above the critical density threshold for condensation, the polariton condensate streams ballistically out of the generation region and around the ring, with a momentum which is quite sensitive to the details of the pump spot location and intensity. For a pump spot near the outer edge of the ring, the condensate was injected with momentum moving toward the inner edge, while for a pump spot near the inner edge, the condensate was injected with outward momentum.

We directly viewed the condensate by imaging the light from the condensate coming through the top mirror. This leakage light means that the condensate has radiative decay, and therefore must be continuously pumped to have a stable state, but it also provides us with a direct probe of the behavior of the condensate. Figures \ref{fig:pos} (b)-(d) show several examples of different patterns in a ring, created by changing the pump location. As seen in these images, the ballistic path of the polaritons wiggles from side to side as it travels around the ring rather than a perfect circular trajectory. The side-to-side motion of the wiggles is affected both by the kinetic energy of the polaritons and the spin-orbit effect, which gives a term completely analogous to the zitterbewegung effect~\cite{z5, z6}. As shown in Figures S7, S8, and S9 of the Supplemental Information \cite{suppl}, the spin-orbit effect is necessary to get good fits of the data; if we leave this term out, the frequency of the wiggles changes substantially.

\begin{figure}
	\centering
	\includegraphics[width=\textwidth]{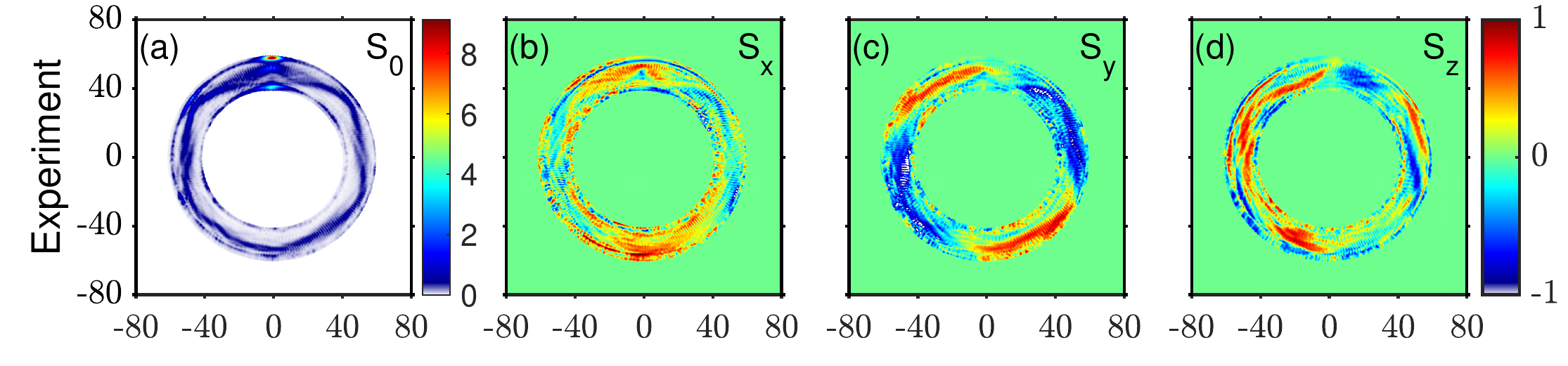}
	\includegraphics[width=\textwidth]{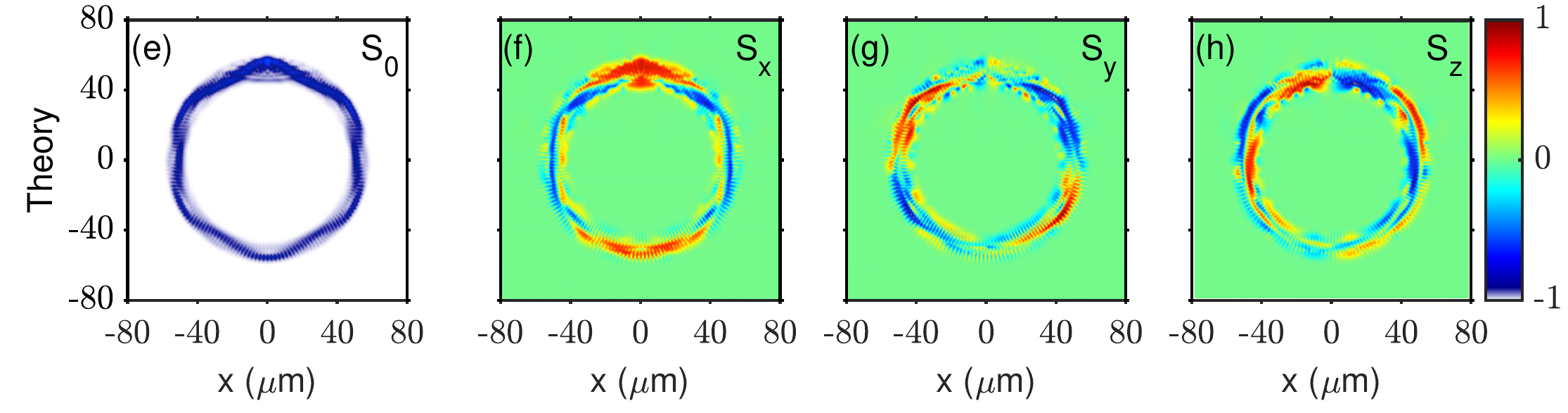}
	\caption{(a) Experimental data of polariton condensate with pump power $6.1\it P_{\rm th}$. The polarization of the pump is 72 degrees. (b)--(d) The three Stokes polarization components. (e)--(h) Theoretical simulation of the condensate and its Stokes components for similar conditions.}
	\label{fig:stokes}
\end{figure}

The exact distance traversed by each wiggle in the ring depends both on the kinetic energy of the ballistic motion, as well as the spin-orbit term discussed below, which gives a ``zitterbewegung'' effect. Figure \ref{fig:stokes} shows both experimental and theoretical images of the polariton condensate motion, showing the total intensity as well as the Stokes polarization components. The theoretical model, discussed below, gives good qualitative agreement with the polarization precession around the ring. The side-to-side ``wiggles'' of the condensate are the result of two effects: the constraint of the quasi-one-dimensional channel of the polaritons and the zitterbewegung effect due to the effective spin-orbit interaction. The Supplementary Information \cite{suppl} for this paper shows that when the zitterbewegung effect is left out of the theory, the side-to-side motion is substantially affected, to the point that the number of wiggles around the ring is changed. 

\begin{figure}
	\centering
	\includegraphics[width=0.9\textwidth]{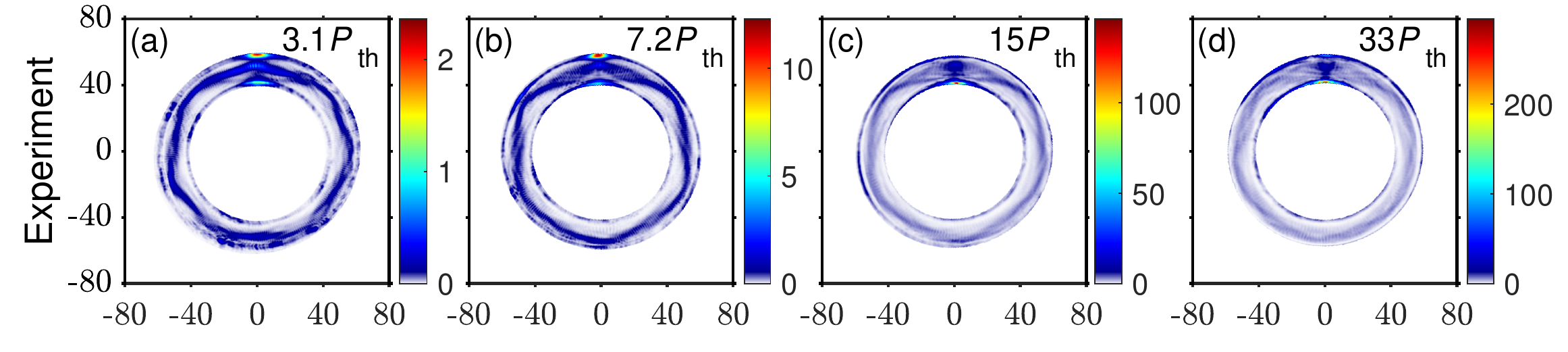}
	\includegraphics[width=0.9\textwidth]{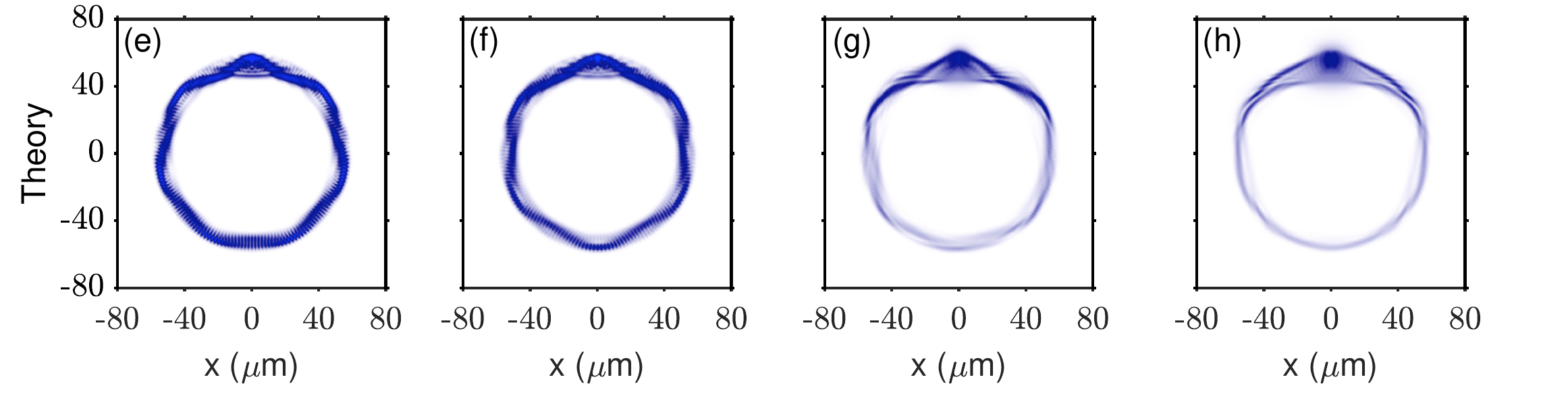}
	\caption{(a)--(d) A series of experimental condensate images as the pump power was increased from $3.3\it P_{\rm th}$ to $32\it P_{\rm th}$. The polarization of the pump for (a)--(d) is 119 degrees, 72 degrees, 111 degrees and 108 degrees, respectively. (e)--(h) The corresponding theoretical simulations. Both experimental data and theoretical simulation show that the polygonal PL didn't change much with low pump power, but ``washed out" at high pump power.}
	\label{fig:pow}
\end{figure}

Figure \ref{fig:pow} shows a series of experimental images with the corresponding images generated by the theoretical model for the motion around a ring as the intensity of the pump is increased. The other experimental conditions remained the same. As seen in this figure, the side-to-side motion is washed out at high density. 

The polarization of the excitation pump is different for each picture, and is recorded in the caption. This is because the pump was at different azimuthal position of the ring, and we rotated the image so that the pump spot is always at the top. However, the polarization of the pump didn’t affect the intensity distribution or Stokes vectors of the condensate, as expected since the non-resonant laser excitation creates hot carriers that must emit many phonons before they turn into polaritons (please see Additional Data of the Experiment section in Supplemental Information \cite{suppl}).

{\bf Theoretical analysis}.
Our theoretical model is based on prior successful work in modeling coherent polariton condensates with generation and decay \cite{ryzhov,ResInOpt4100105,PhysRevLett126075302,NewJPhys22083059,PRB595082}.
In the linear conservative limit, polaritons in the considered system obey the Hamiltonian
\begin{equation}
\label{EqH0}
\hat{H}_0 = \frac{\hbar^2 \hat{k}^2 }{2 m^*} \hat{\sigma}_0 + V _{\text{st}} (\mathbf{r}) \hat{\sigma}_0 + \hbar \hat{ \boldsymbol{\Omega}} \cdot \hat{\mathbf{S}},
\end{equation}
where $m^*$ is the effective mass of polaritons, $\hat{\mathbf{k}} = (\hat{k}_x, \hat{k}_y) = (-\mathrm{i} \partial _x,-\mathrm{i} \partial _y)$ is the quasimomentum operator.
$V _{\text{st}} (\mathbf{r})$ describes the ring-shaped stationary potential.
The Hamiltonian acts on the polariton quantum state described by the spinor $| \Psi \rangle = [\Psi _+ (\mathbf{r}), \Psi _- (\mathbf{r})]^{\mathrm{T}}$,  where $ \Psi _{\pm} (\mathbf{r}) $ are the wave functions of the spin up and spin down states.
The last term in~\eqref{EqH0} is responsible for the spin-orbit interaction (SOI) of polaritons originated from the splitting in the TE- and TM-polarized polariton modes in the layered microcavity structure~\cite{PRB595082, physsolidstate1999}.
The effect of SOI can be described as precession of the polariton pseudospin vector $\hat{\mathbf{S}} = (\hat{S}_x, \hat{S}_y, \hat{S}_z)$ around the effective magnetic field $\hat{\mathbf{\Omega}}$ induced by the splitting:
\begin{equation}
\label{EqEffField}
\hat{\boldsymbol{\Omega}} = [\Delta (\hat{k}_x^2 - \hat{k}_y^2),2 \Delta \hat{k}_x \hat{k}_y,0],
\end{equation}
where $\Delta$ is the TE-TM splitting constant.
The pseudospin vector is introduced as $\hat{\mathbf{S}} = \frac{1}{2} \hat{\boldsymbol{\sigma}}$, where $\hat{\boldsymbol{\sigma}}$ is the vector of the Pauli matrices, $\hat{\sigma}_0$ is the $2 \times 2$ identity matrix.

The trajectory of a ballistically propagating particle is described by the equation of motion for the position operator: $\hbar d _t \hat{r}_j = \mathrm{i} [\hat{H}_0, \hat{r}_j]$, where $\hat{r}_j=\hat{x}, \hat{y}$.

The structure considered here supports TE-TM splitting, which significantly modifies the waveguide modes.
The SOI component of the Hamiltonian~\eqref{EqH0} introduces an oscillating term to the equation of motion of the position operator~$\hat{\mathbf{r}}$,
which is supplemented with the equation of precession of the polariton pseudospin $d _t\hat{\mathbf{S}} = \hat{\boldsymbol{\Omega}} \times \hat{\mathbf{S}}$.

To underpin the discussed experimental observations, we performed numerical simulations of the behavior of the spinor exciton-polariton condensate in a ring waveguide trap.
We use the generalized Pauli equation for the spinor $| \Psi \rangle = [\Psi _+ (t,\mathbf{r}), \Psi _- (t,\mathbf{r})]^{\mathrm{T}} $, where $ \Psi _{\pm} (t,\mathbf{r}) $ are the wave functions of the spin up and spin down states.
See details of the model in the Supplemental Information \cite{suppl} (also references \cite{suppl_1, suppl_2} therein).
To get closer to the experimental conditions, we take into account the effects of interaction of particles and the non-conservative processes of the pump that generates the polaritons and radiative losses.

Polaritons in the system possess a finite lifetime due to escaping photons from the structure through the top mirrors and due to additional damping induced by etching of the microcavity.
The latter includes tunneling of photons through the side walls of the ring trap and scattering of polaritons from the rough surface of the walls.
This effect can be taken into account by assuming the stationary potential $V _{\text{st}} (\mathbf{r})$ to be complex~\cite{ResInOpt4100105,PhysRevLett126075302}.
The non-resonant optical pump excites the spinor reservoir of incoherent excitons (in fact, exciton-like polaritons) which plays two important roles in the formation of the polariton intensity patterns. 
First, it feeds the polariton condensate via stimulated relaxation of particles from it.
Second, it contributes to the effective potential landscape for polaritons due to the repulsive interaction of particles in the condensate and the reservoir.
The cloud of excitons forms the local potential maximum under the pump spot which height increases with increasing pump power.
Polaritons emerging within the pump spot propagate away from the potential maximum; as they leave, their potential energy is converted to kinetic energy of ballistic motion.
As one can see from~\eqref{EqEffField}, the Larmor frequency of the effective magnetic field depends quadratically on the polariton wave vector.
The pump power together with the displacement of the pump spot play the role of the control parameters in our system. The density dependence shown in Figure \ref{fig:pow} is modeled simply by turning up the pump power, which increases the role of the nonlinear polariton-polariton and polariton-exciton terms. 

\begin{figure}
\includegraphics[scale = 0.7]{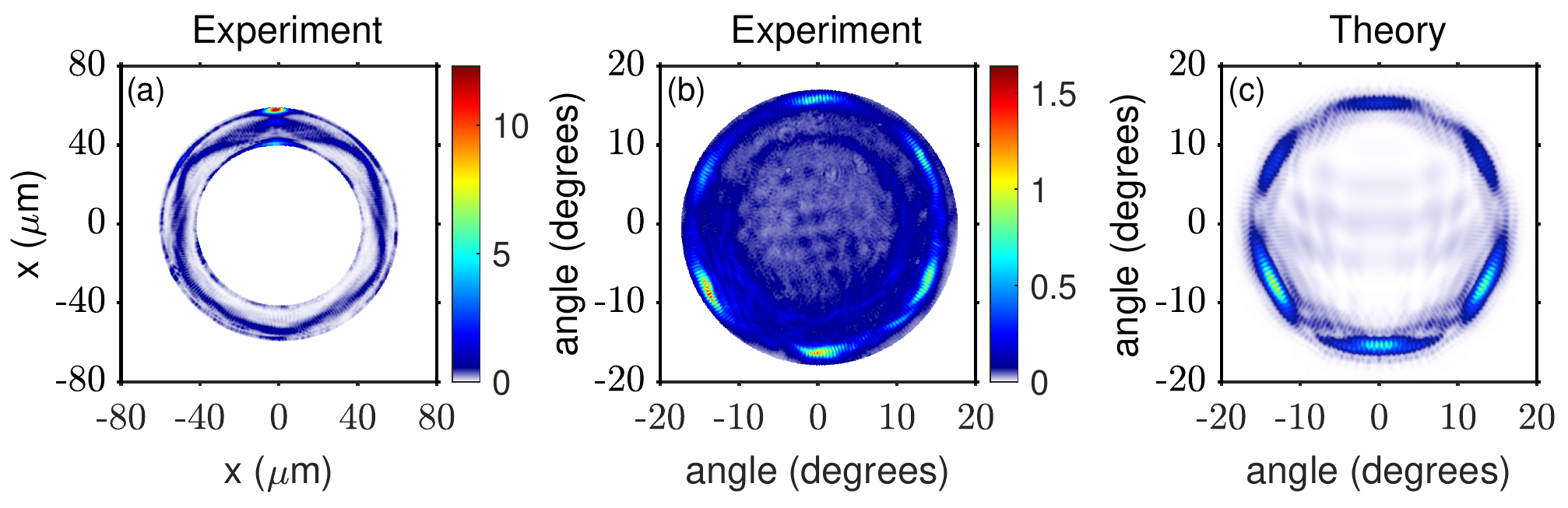}
\includegraphics[scale = 0.7]{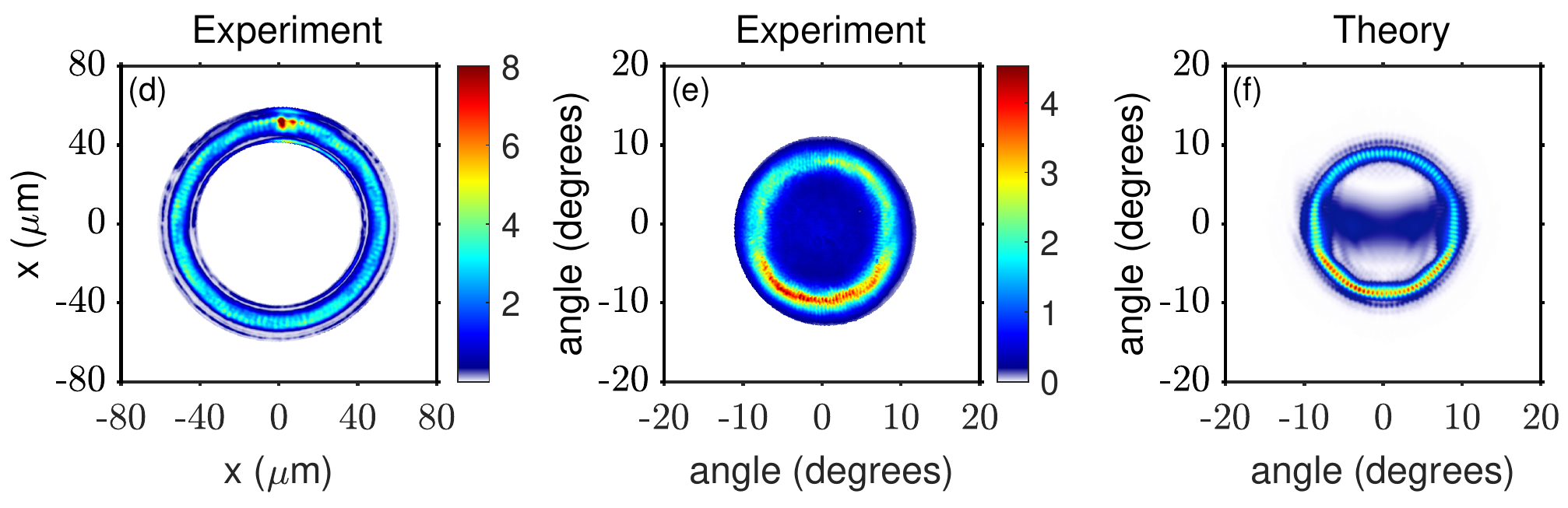}
\caption{(a) and (d): real-space images showing two different patterns due to different radial pumping position. The pump power was slightly differet, $6.1P_{th}$ for (a), and  $5.3P_{th}$ for (d).  (b) and (e): the corresponding experimentally measured momentum-space images of (a) and (d). (c) and (f): theoretical momentum-space images for (b) and (e).}
\label{fig:k}
\end{figure}

Finally, Fig.~\ref{fig:k} shows the momentum-space images for both the experimental data (done by recording the Fourier plane of the polariton light emission) and the theory (done by taking the Fourier transform of the solutions of the theoretical model). For purely circular motion in the ring, the $k$-space distribution will also be a ring. Because the orbitals have favored angles, the $k$-space distribution is seen as bright spots at specific directions.

{\bf Conclusions}. In these experiments, we have shown that we can create stable orbital-like patterns by varying the position of the injection spot, which in turn controls the ballistic injection momentum of the polariton condensate. Our polarization resolution shows that the effective spin-orbit coupling plays a major role in controlling the precession of the polarization around the ring, as well as the effective length of the path for the phase advance. Further, the theoretical analysis shows that the zitterbewegung effect due to the effective spin-orbit coupling plays a substantial role in the orbital pattern of the ballistic condensate.

Bose condensation is crucial for this effect to be seen, because the condensation locks the particles into mono-energetic states with a single ballistic behavior. Without this, the particles would occupy a continuum of modes that would obscure each other in time-integrated imaging.

{\bf Acknowledgements}. The experimental work at Pittsburgh and sample fabrication at Princeton was funded by the National Science Foundation (Grant No. DMR-2004570). 
Numerical simulations were supported by Saint Petersburg State University (Grant No. 91182694).
A.K.~acknowledges support from the Moscow Institute of Physics and Technology under the Priority 2030 Strategic Academic Leadership Program.

\end{document}


\renewcommand*{\thefigure}{S\arabic{figure}}
\renewcommand*{\thesection}{S\arabic{section}}
\renewcommand*{\theequation}{S\arabic{equation}}

\title{Supplemental information: Ballistic transport of a polariton ring condensate with spin precession}

\author{Q. Yao$^1$, E. Sedov$^{2,3,4,5}$, S. Mukherjee,$^6$ J. Beaumariage$^1$, B. Ozden,$^7$ K. West,$^8$ L. Pfeiffer,$^8$\\
A. Kavokin,$^{2,3,4,5,9}$ and  D. W. Snoke$^1$}

\affiliation{$^1$Department of Physics and Astronomy, University of Pittsburgh\\ 3941 O'Hara St., Pittsburgh, PA 15260, USA\\
$^2$Key Laboratory for Quantum Materials of Zhejiang Province, School of Science, Westlake University, 18 Shilongshan Rd, Hangzhou 310024, Zhejiang, China\\
$^3$Westlake Institute for Advanced Study, 18 Shilongshan Rd, Hangzhou 310024, Zhejiang, China\\
$^4$Spin Optics Laboratory, St. Petersburg State University, Ulyanovskaya 1, St. Petersburg 198504, Russia\\
$^5$Russian Quantum Center, 100 Novaya St., Skolkovo, Moscow region, 143025, Russia\\
$^6$Joint Quantum Institute, University of Maryland and National Institute of Standards and Technology, College Park,
Maryland 20742, USA\\
$^7$Department of Physics, Penn State Abington, Abington, PA 19001, USA\\
$^8$Department of Electrical Engineering, Princeton University, Princeton, NJ 08544, USA \\
$^9$Moscow Institute of Physics and Technology, Dolgoprudnyi, Moscow Region, 141701, Russia
}

\maketitle
\section{Experimental Setup}
\begin{figure}[h!]
    \centering
    \includegraphics[scale=0.35]{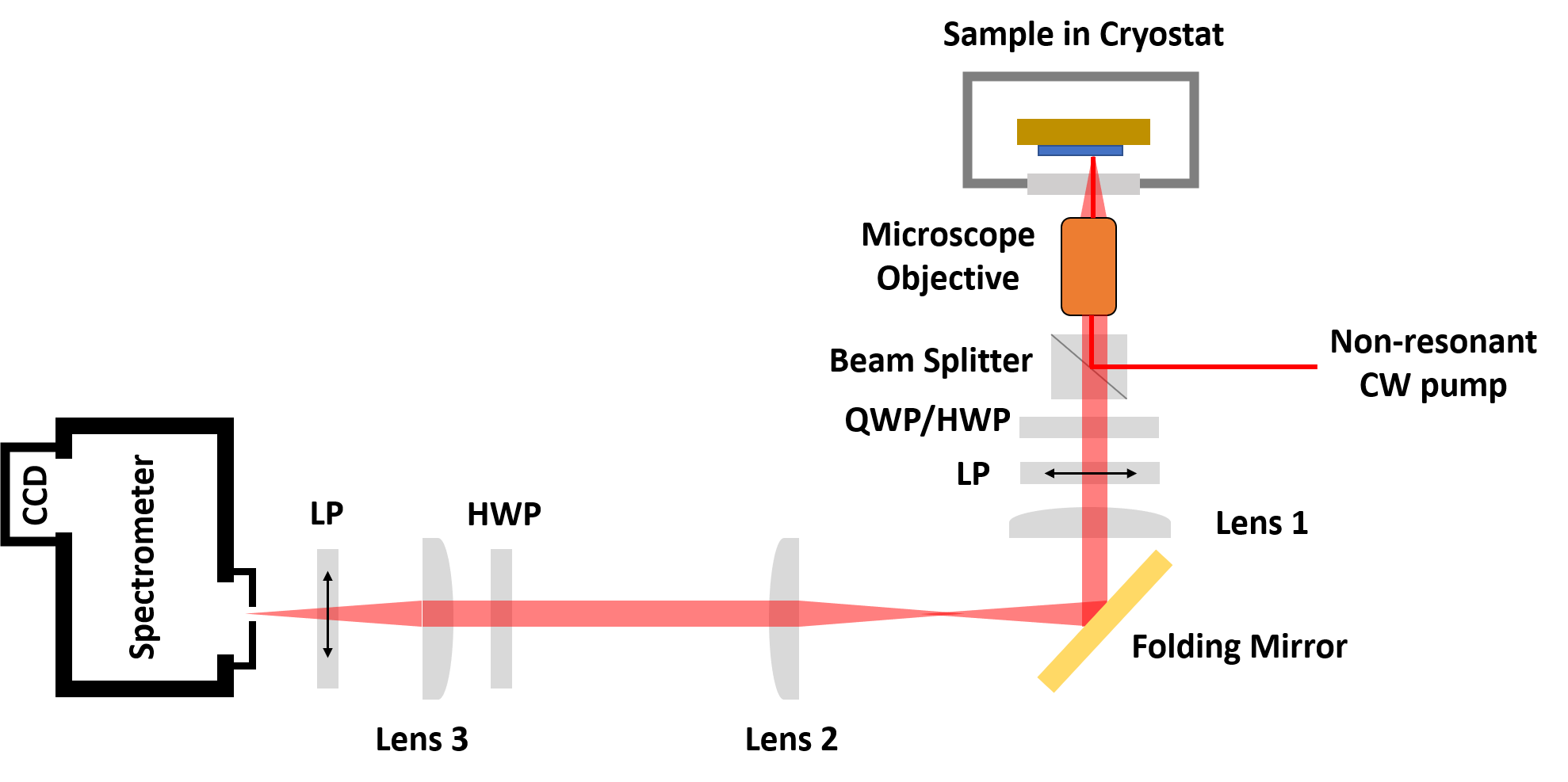}
    \caption{Schematic of the experimental setup. QWP: quarter waveplate, HWP: half waveplate, LP:linear polarizer}
    \label{fig:schematic}
\end{figure}

We followed the method of Ref.~\cite{shouvik} for measuring the full set of components of the Stokes polarization vector. Fig.~\ref{fig:schematic} is a schematic of the experimental setup. The polarization information of the PL is obtained by adding the half/quarter waveplate and the linear polarizer just after the cube beam splitter. Table.~\ref{tab:combination} gives the combination of the waveplate and polarizer and corresponding information. $x$ and $y$ are horizontal and vertical directions in the lab frame, respectively. $\theta_{\lambda/2}$ and $\theta_{\lambda/4}$ are angles between vertical (hence $y$) direction and fast axis of half/quarter waveplate. The transmission axis of linear polarizer was kept vertical during the experiment. $E_x$ and $E_y$ are complex electric field components of polarized PL, and $\epsilon$ is electric field of non-polarized PL. 

\begin{table}[b]
    \centering
    \begin{tabular}{ccccc}
    \hline
    \hline
        $\theta_{\lambda/2}$ & $\theta_{\lambda/4}$ & Polarizer & Measured Intensity & Observed Intensity \\
        \hline
        0 & - & vertical & $|E_y|^2+\epsilon^2/2$ & $I_1$ \\
        
        $\pi/8$ & - & vertical & $\frac{1}{2}(|E_x|^2+|E_y|^2+(E_x^*E_y+E_y^*E_x)+\epsilon^2)$ & $I_2$ \\
        
        $\pi/4$ & - & vertical & $|E_x|^2+\epsilon^2/2$ & $I_3$ \\
        
        - & $\pi/4$ & vertical & $\frac{1}{2}(|E_x|^2+|E_y|^2+i(E_y^*E_x-E_x^*E_y)+\epsilon^2)$ & $I_4$ \\
        
        - & $7\pi/4$ & vertical & $\frac{1}{2}(|E_x|^2+|E_y|^2-i(E_y^*E_x-E_x^*E_y)+\epsilon^2)$ & $I_5$ \\
        \hline
    \end{tabular}
    \caption{} 
    \label{tab:combination}
\end{table}

Although the signal right after the first linear polarizer was vertically polarized, it would change polarization when reaching the slit after going through all the optics. Therefore, anther half waveplate and polarizer were placed close to the slit. The transmission axis of the second polarizer was also kept vertical during the experiment, which removed the polarization sensitivity of the spectrometer and camera. For each measurement, two images were taken, one with the half waveplate (the one close to slit) fast axis at $0^{\circ}$, and the other with it at $45^{\circ}$. By adding the two images together, we got the total intensity in each the measurement $I_1 \sim I_5$, free from the polarization sensitivity of the system.

\begin{figure}[t]
    \centering
    \includegraphics[scale=0.5]{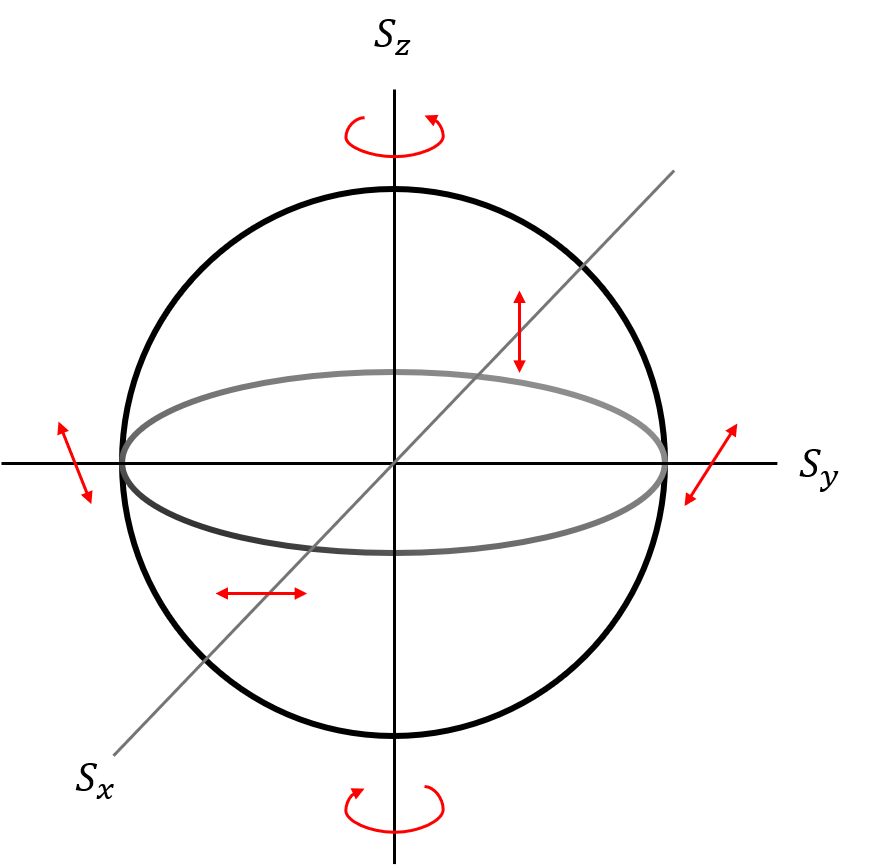}
    \caption{Definition of Stokes Vectors on the Bloch sphere}
    \label{fig:stokesDef}
\end{figure}

From the observed intensity $I_1 \sim I_5$, we can obtain Stokes vectors $(S_0, S_x, S_y, S_z)$:

\begin{equation}
    \begin{aligned}
    &S_0 = I_{tot} = |E_x|^2+|E_y|^2+\epsilon^2 = I_1+I_3 = I_4+I_5\\
    &S_x = \frac{|E_x|^2-|E_y|^2}{|E_x|^2+|E_y|^2+\epsilon^2} = \frac{I_3-I_1}{I_1+I_3} \\
    &S_y = \frac{E_x^*E_y+E_y^*E_x}{|E_x|^2+|E_y|^2+\epsilon^2} = \frac{2I_2-I_1-I_3}{I_1+I_3} \\
    &S_z = \frac{i(E_x^*E_y-E_y^*E_x)}{|E_x|^2+|E_y|^2+\epsilon^2} = \frac{I_5-I_4}{I_4+I_5}
    \end{aligned}
\end{equation}
The definition of Stokes vectors on Bloch sphere is shown in Fig. \ref{fig:stokesDef}

\newpage
\section{Characteristics of the Ring}
\begin{figure}[h!]
    \centering
    \includegraphics[scale=0.6]{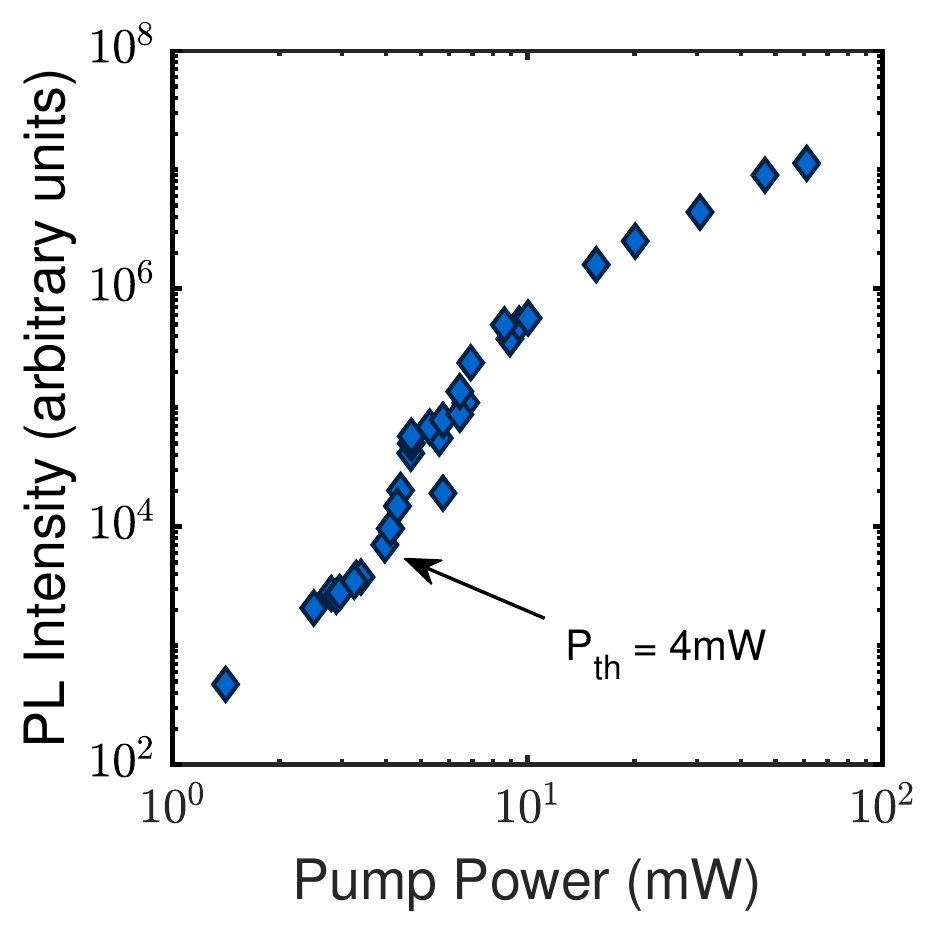}
    \caption{Power Series}
    \label{fig:powerSeries}
\end{figure}

Fig. \ref{fig:powerSeries} shows the PL intensity of one position on the ring versus pump power when focusing the laser on the other side of the ring. It indicates the threshold power was about 4 mW.

\begin{figure}[t!]
    \centering
    \includegraphics[scale=0.6]{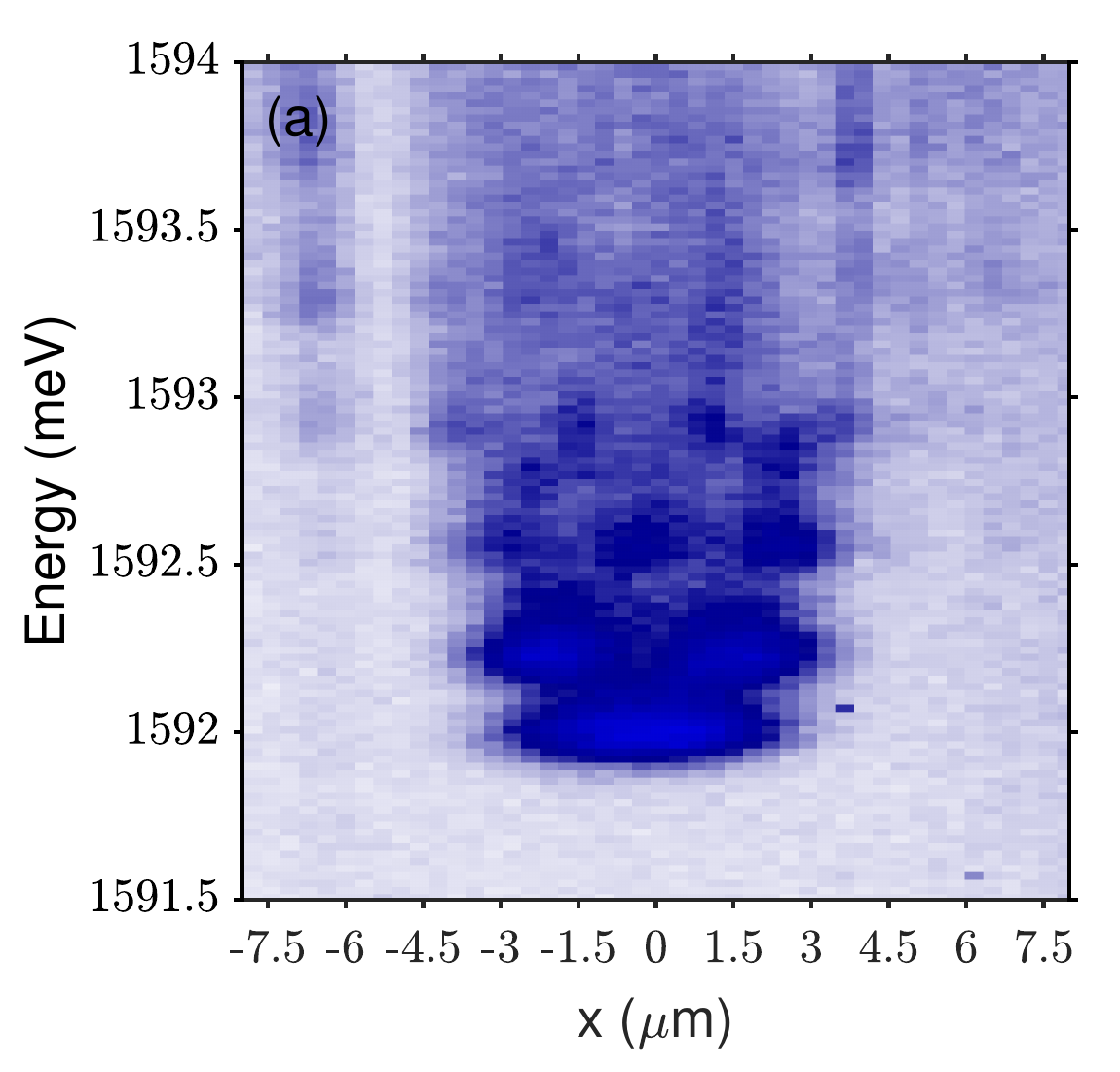}
    \includegraphics[scale = 0.6]{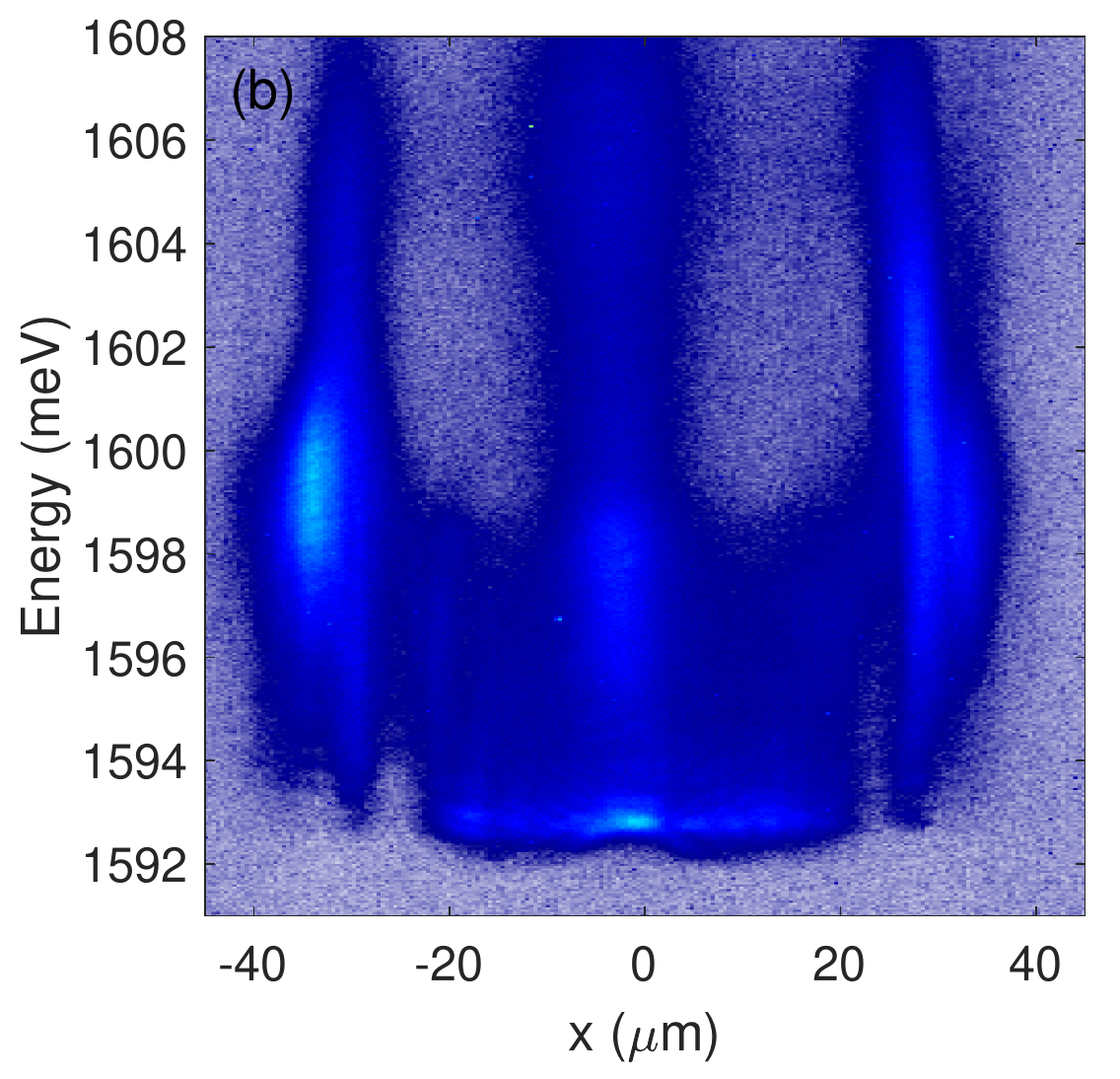}
    \includegraphics[scale = 0.6]{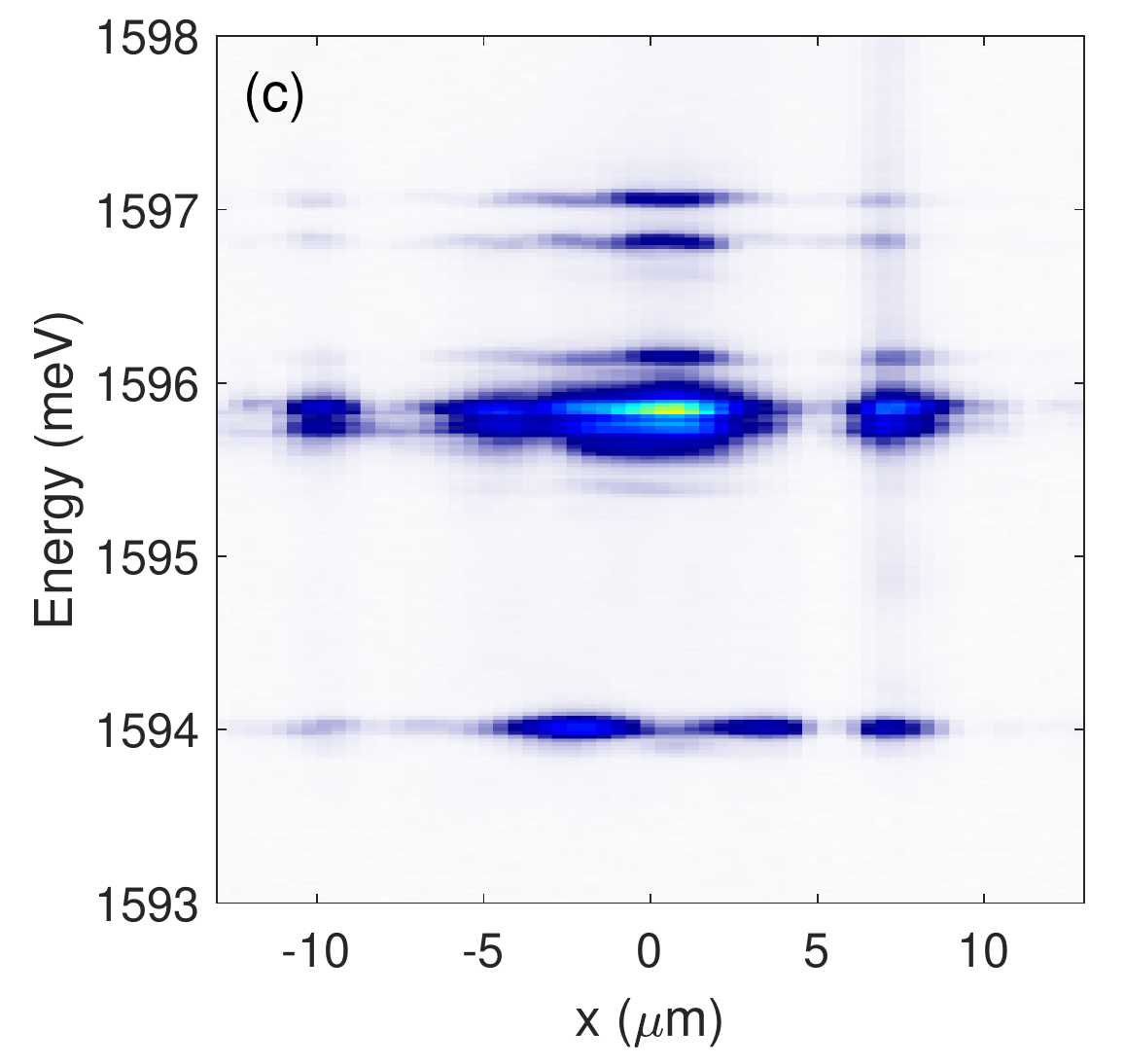}
    \caption{(a) Energy-resolved real-space image across the ring when pumping at low power. Quantized states for radial motion separated by $200~\mu$eV were observed. (b) Energy hill at the pump spot ($x = 0\mu$m). The pump power was $P = 0.62P_{th}$, and the real-space slit was tangential to the ring. Signal near $x=\pm 35\mu$m was from quantum wells outside the ring. (c) Energy-resolved real-space image in condensate regime. The data corresponds to PL pattern in Fig. 4(a). The slit was along radial direction. Only one state is dominant.}
    \label{fig:EvRlowPower}
\end{figure}

Fig. \ref{fig:EvRlowPower}(a) shows the quantized states due to the confinement in radial direction at low pump power. The energy separation was about $200~\mu$eV. Fig. \ref{fig:EvRlowPower}(b) shows the energy hill at the pump spot location ($x = 0\mu$m) with pump power $P = 0.62P_{th}$. The real-space slit was tangential to the ring. The signal near $x =\pm35\mu$m was from quantum wells outside the ring. 

\newpage

\section{Additional data of the experiment}
\begin{figure}[h]
    \centering
    \includegraphics[scale=0.6]{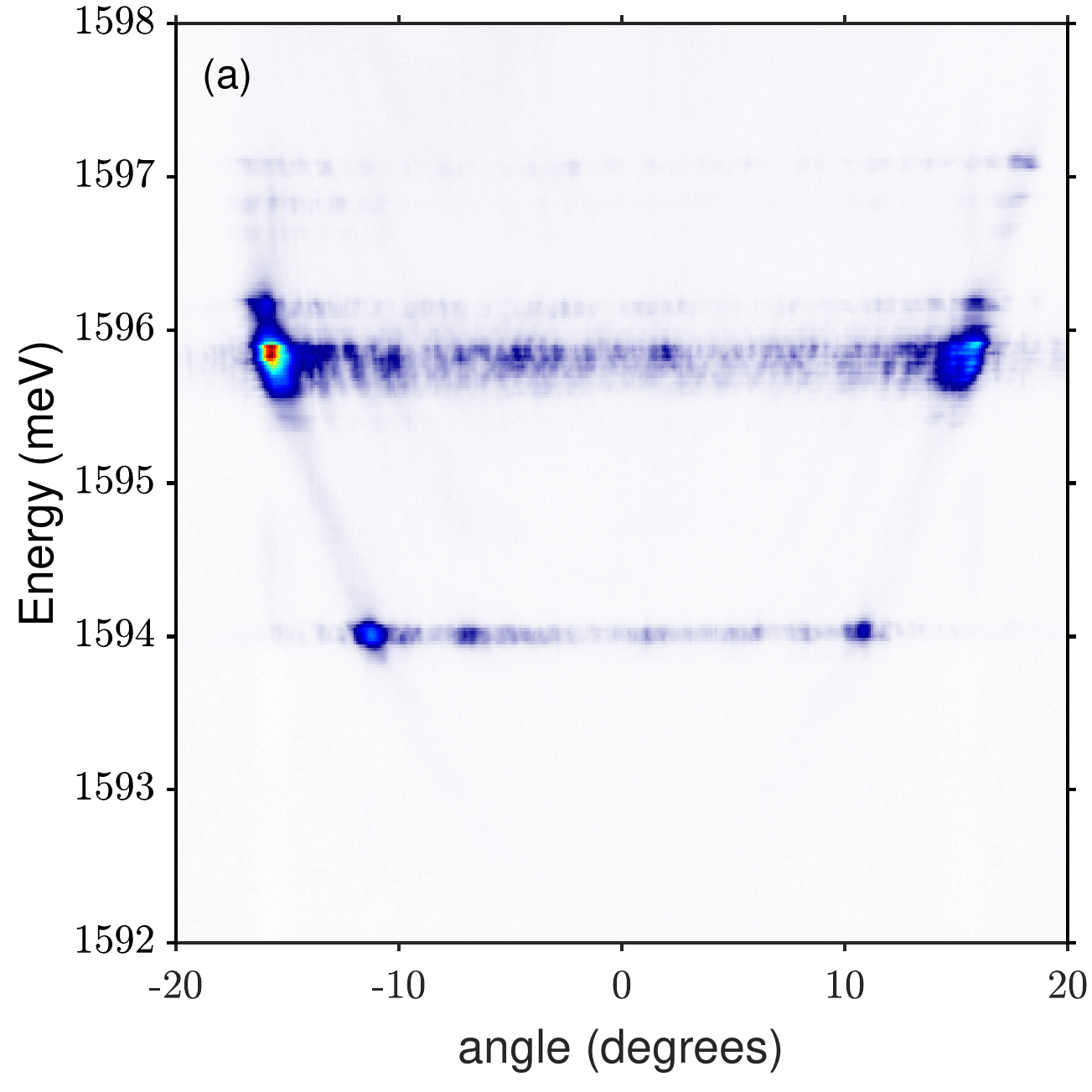}
    \includegraphics[scale=0.6]{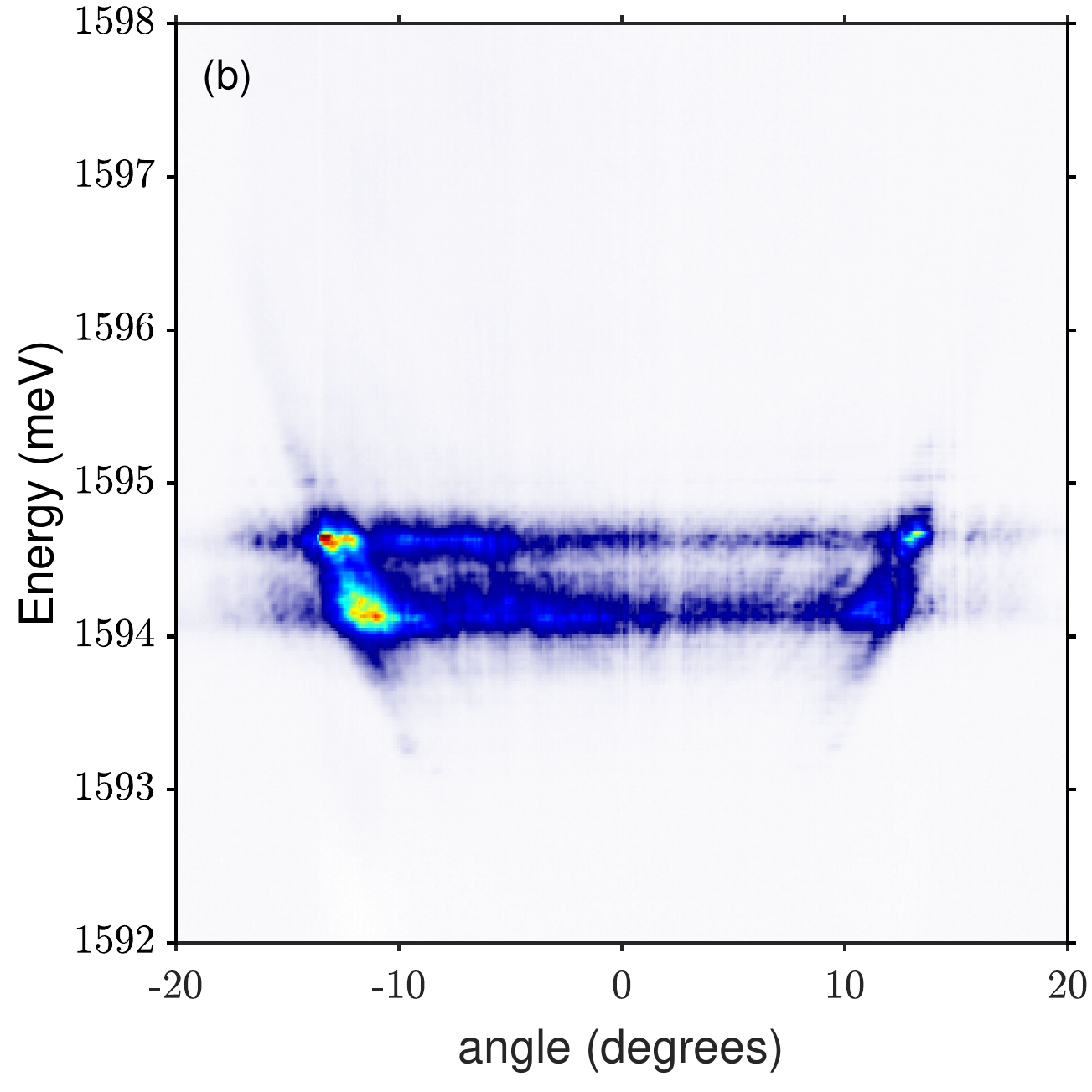}
    \caption{Energy-resolved momentum-space images of the ring. (a) corresponds to PL pattern in Fig.4(a) in the main text. It was taken by energy resolving the central slit of Fig.4(b). (b) corresponds to PL pattern similar to Fig.4(d). }
    \label{fig:EvK}
\end{figure}
Fig. \ref{fig:EvK} shows the energy-resolved momentum-space images for two different PL patterns. Fig. \ref{fig:EvK}(a) corresponds to pattern in Fig. 4(a) in the main text and it was taken by energy resolving the central slit of Fig. 4(b). Fig. \ref{fig:EvK}(b) corresponds to pattern similar to Fig. 4(d). As seen here, the condensate can have multiple ballistic modes, but is usually dominated by one mode. This was also observed in energy-resolved real-space image (Fig. \ref{fig:EvRlowPower}(c)).

\begin{figure}[t!]
    \centering
    \includegraphics[scale=0.6]{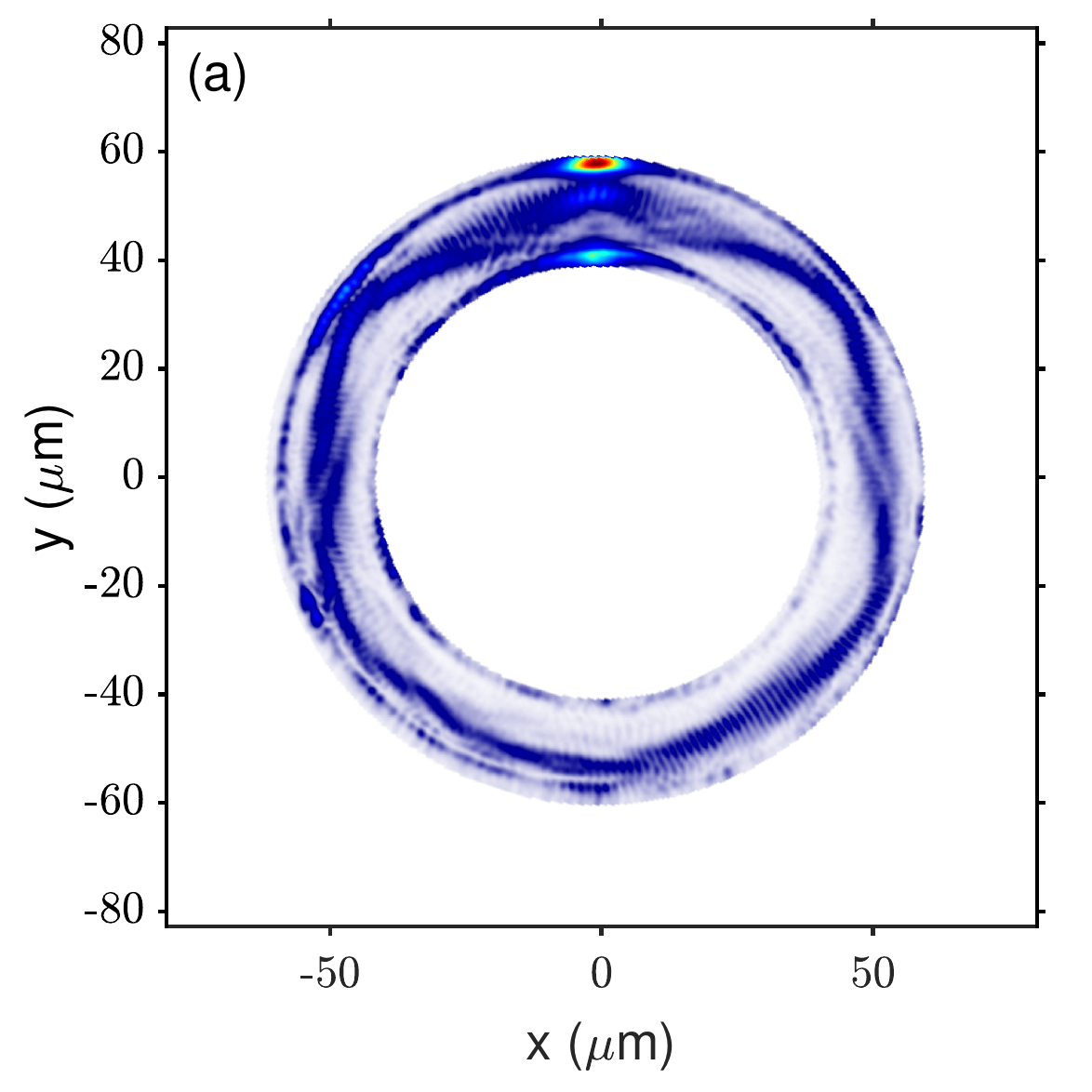}
    \includegraphics[scale=0.6]{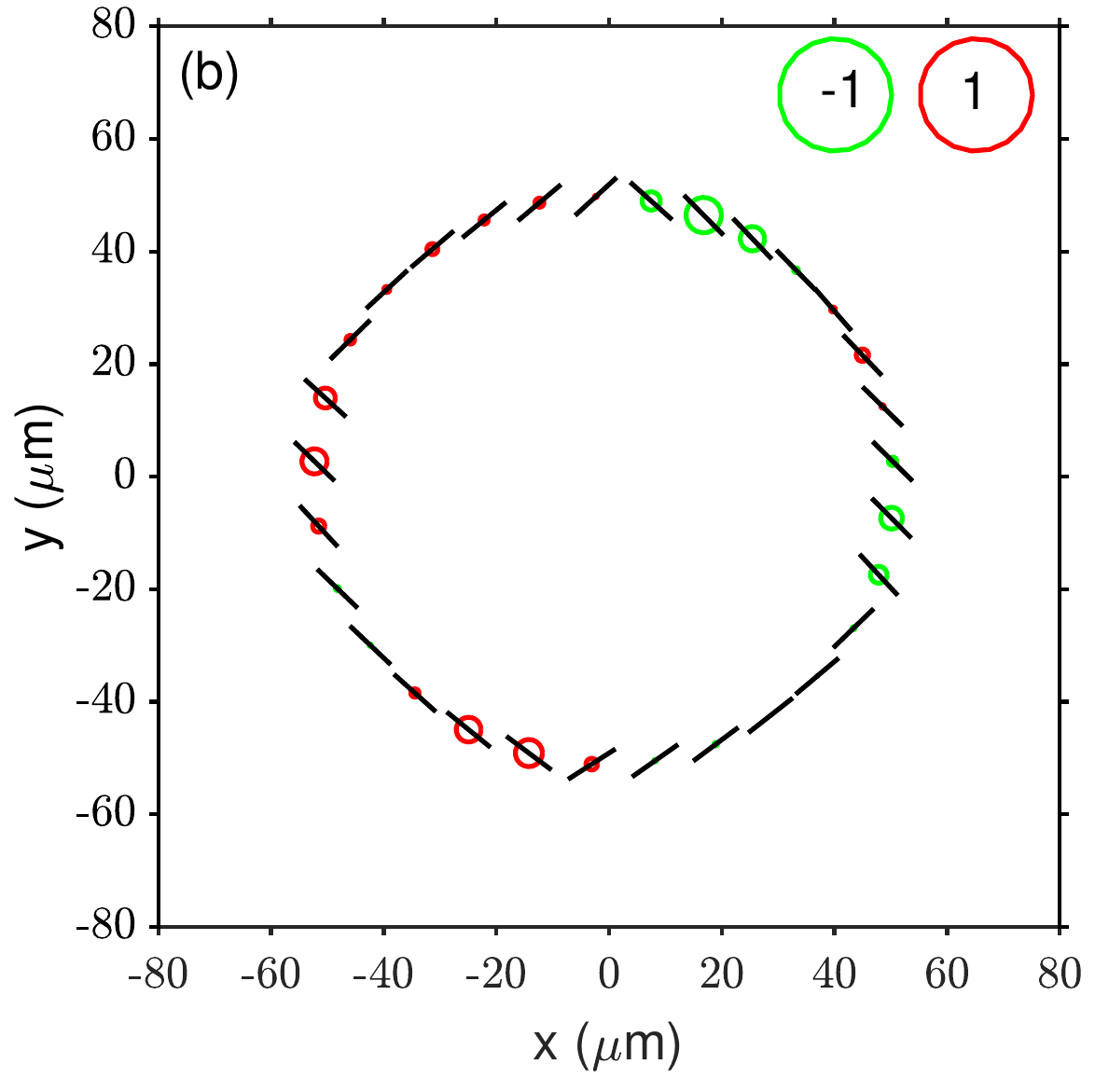}
    \includegraphics[scale=0.6]{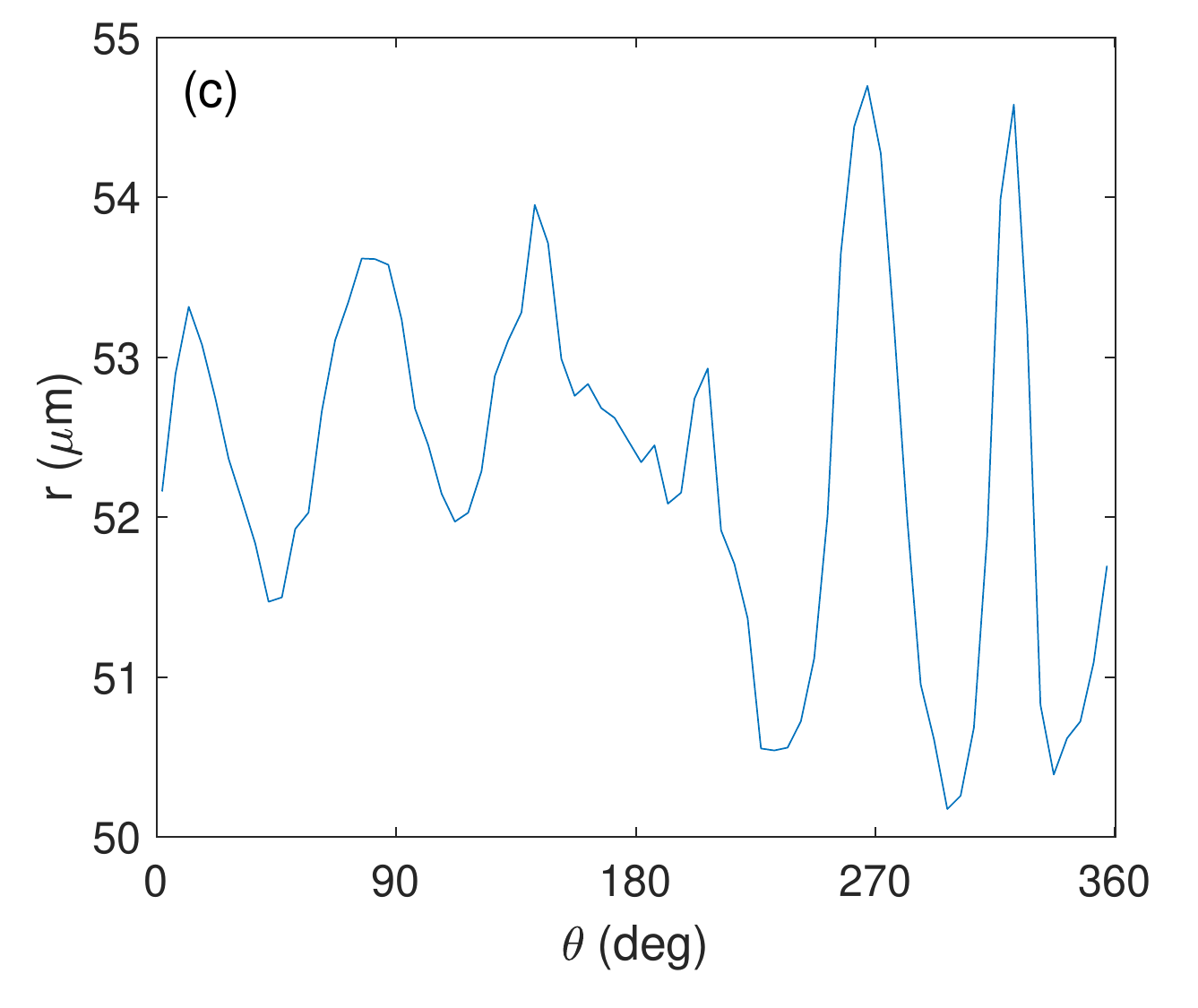}
    \caption{(a) Real-space image shown in Fig. 4(a) in the main text. (b) Polarization map of the experimentally measured polarization state. Black arrows show the linear polarization direction. Circles represent circular polarization (red left-handed, green right-handed). The radii of the circle is proportional to the degree of circular polarization. (c) radial center-of-mass coordinate on the polar angle.}
    \label{fig:polMap}
\end{figure}

Fig. \ref{fig:polMap}(b) is polarization map constructed by using Stokes vectors, Fig. 2(b)-(d) in the main text. It clearly shows that the linear polarization is precessing along the ring due to spin-orbit interaction. Fig. \ref{fig:polMap}(c) is radial center-of-mass coordinate on the polar angle extracted from Fig. \ref{fig:polMap}(a).

\begin{figure}[h!]
    \centering
    \includegraphics[scale=0.75]{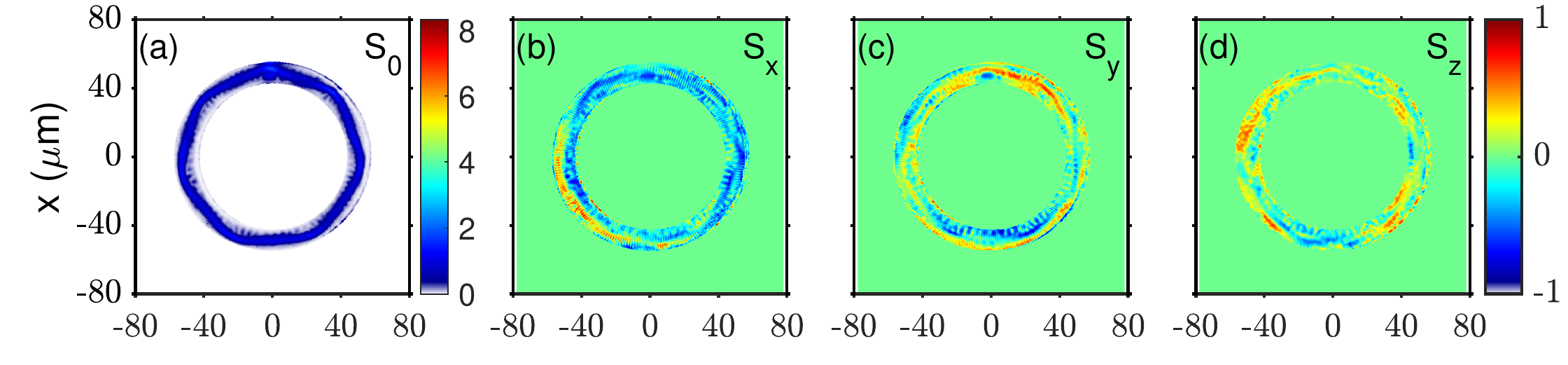}
    \includegraphics[scale=0.75]{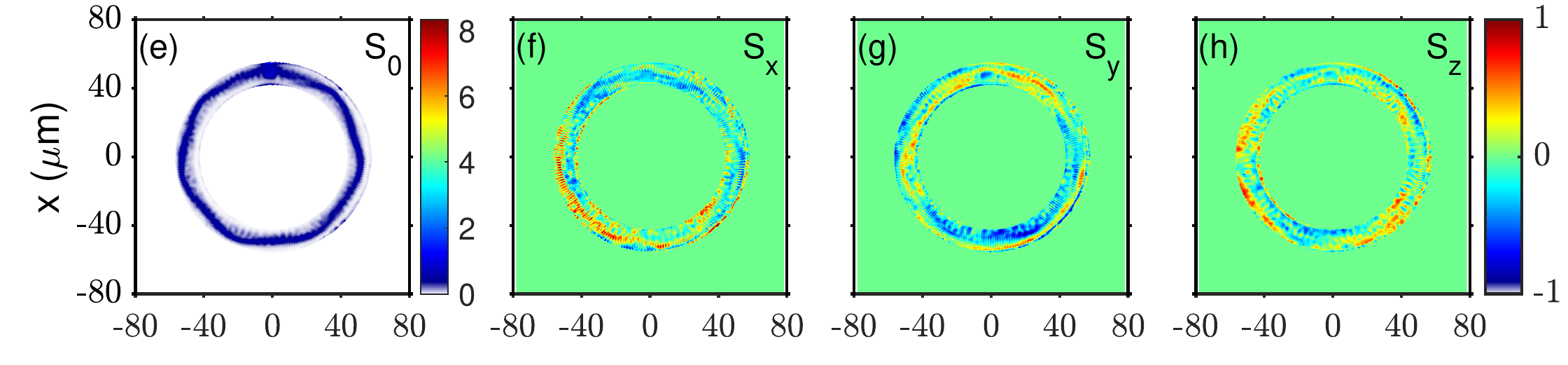}
    \includegraphics[scale=0.75]{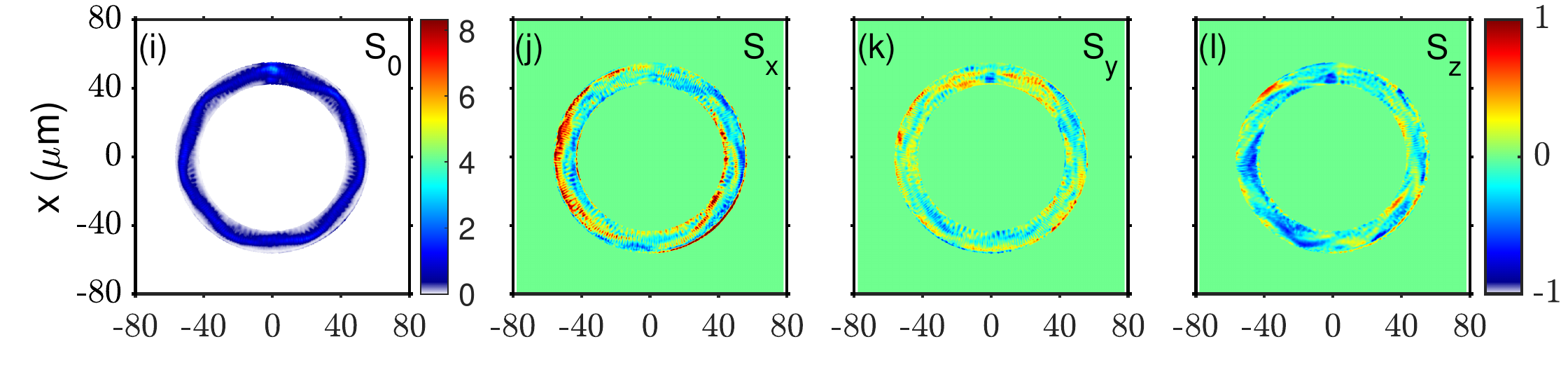}
    \includegraphics[scale=0.75]{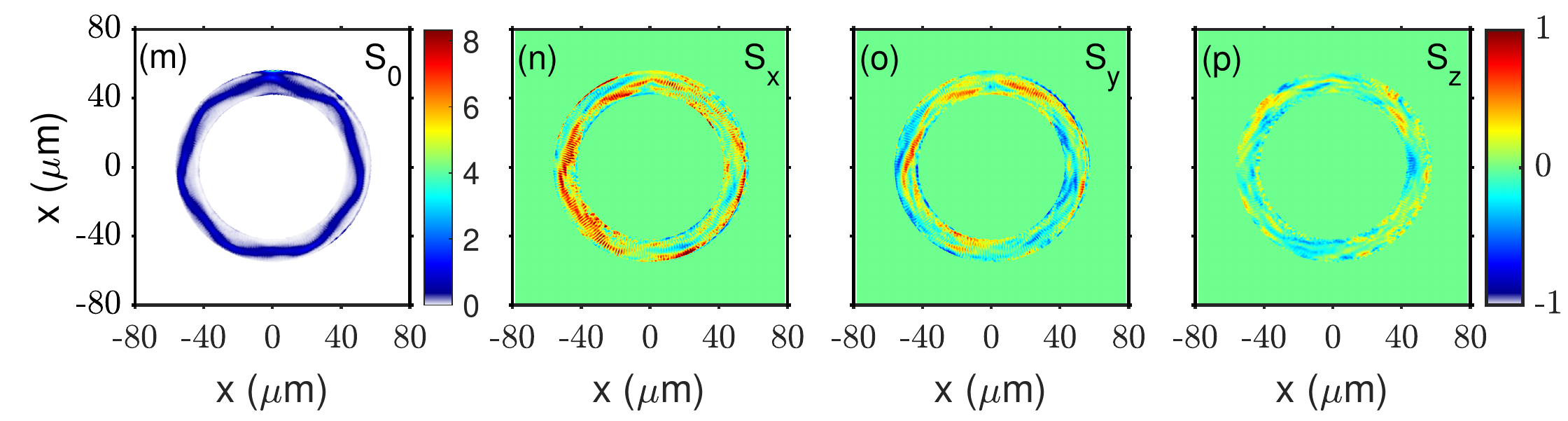}
    \caption{Real-space image and Stokes vectors when the pump is (a)-(d) closed to horizontally polarized (-8 degrees), (e)-(h) close to vertically polarized (82 degrees), (i)-(l) close to $45$ deg polarized (37 degrees), (m)-(p) close to $-45$ deg polarized (-53 degrees).}
    \label{fig:Spol}
\end{figure}

\newpage
Photons from the non-resonant pump enter the microcavity through the reflectivity dips and generate high-energy excitons. Then the high-energy excitons scatter to lower polaritons states at $k=0$ via phonon-exciton interaction and polariton -polariton interaction. In the process of interacting with lattice (phonons), the polarization information of the original photon is lost. Our pump laser is horizontally polarized in the lab frame. However, the polarization of the pump is different for the data shown in the main text because the pump was at different azimuthal position of the ring in the lab frame and we rotated the image so that the pump spot is always at the top but the pump is no longer horizontally polarized relative to the paper. In order to verify that the pump polarization doesn't affect the polarization of the condensate, I controlled the polarization of the pump while maintaining its position and the power. Fig. \ref{fig:Spol} shows the results when the pump was linear polarized and the polarization directions were in four directions. The Stokes vectors in those cases look almost the same. 

\newpage
\section{\label{SSecModel} Numerical model}

We model the behavior of the polariton condensate using the generalized Pauli equation for the spinor $| \Psi \rangle = [\Psi _+ (t,\mathbf{r}), \Psi _- (t,\mathbf{r})]^{\mathrm{T}} $, where $ \Psi _{\pm} (t,\mathbf{r}) $ are the wave functions of the spin up and spin down states:
\begin{equation}
\label{SEqGGPE}
\mathrm{i} \hbar \partial _t |\Psi \rangle = \left[ \hat{H}_0  + \hat {V} _{\text{eff}} \right] | \Psi \rangle + \frac{\mathrm{i} \hbar }{2} (\hat{R}_{\text{in}} - \gamma \hat{\sigma}_0) | \Psi \rangle
\end{equation}
coupled to the rate equation for the reservoir of incoherent excitons characterized by the vector $| n _{\mathrm{R} }\rangle = [n_{\mathrm{R}+} (t,\mathbf{r}), n _{\mathrm{R}-}  (t,\mathbf{r})]^{\mathrm{T}} $, where $n _{\mathrm{R}\pm}  (t,\mathbf{r})$ are densities of bright excitons with spins up and down:
\begin{equation}
\label{SEqResRateEq}
\partial _t |n _{\mathrm{R}} \rangle = | P \rangle - (\gamma _{\mathrm{R}} \hat{\sigma}_0 + \hat{R} _{\text{out}}) | n_{\mathrm{R}} \rangle.
\end{equation}

The Hamiltonian $\hat{H}_0$ in~\eqref{SEqGGPE} describes the polariton energy including the spin-orbit interaction (SOI) in the linear regime:
\begin{equation}
\label{SEqH0}
\hat{H}_0 = \frac{\hbar ^2 \hat{k}^2}{2 m^*} \hat{\sigma}_0 
+ V_{\text{st}} (\mathbf{r}) \hat {\sigma}_0
+ \hbar \hat{\boldsymbol{\Omega}} \cdot \hat{\mathbf{S}},
\end{equation}
where $m^*$ is the polariton effective mass, $\hat{\mathbf{k}} = (\hat{k}_x, \hat{k}_y) = (-\mathrm{i} \partial _x,-\mathrm{i} \partial _y)$ is the quasimomentum operator.
We consider the TE-TM splitting as the origin of SOI.
We introduce the polariton pseudospin operator $\hat{\mathbf{S}} = \frac{1}{2} \hat{\boldsymbol{\sigma}}$, where $\hat{\boldsymbol{\sigma}}$ is the vector of Pauli matrices, $\hat{\sigma}_0$ is the $2\times2$ identity matrix.
Within the pseudospin formalism, SOI is described as the effective magnetic field which induces precession of the polariton pseudospin.
The effective magnetic field operator $\hat{\boldsymbol{\Omega}}$ is given as follows: 
\begin{equation}
\label{SEqEffField}
\hat{\boldsymbol{\Omega}} = [\Delta (\hat{k}_x^2 - \hat{k}_y^2),2 \Delta \hat{k}_x \hat{k}_y,0],
\end{equation}
where $\Delta$ is the TE-TM splitting constant.

Details of spin relaxation of exciton-polariton in microcavities can be found in Ref.~\cite{kavokin}. In brief, Ref.~\cite{maialle} points that the longitudinal-transverse splitting (LT-splitting) of excitons in quantum wells is a result of the long-range exchange interaction, which can be described by the reduced spin Hamiltonian of Pikus-Bir as 

\begin{equation}\label{LT}
H_{ex} \sim \Delta_{LT}
\begin{bmatrix}
k^2 & (k_x-ik_y)^2 \\
(k_x+ik_y)^2 & k^2
\end{bmatrix}
\end{equation}
where $\Delta_{LT}$ is the longitudinal-transverse splitting of a bulk exciton. Off-diagonal terms of the Hamiltonian~\ref{LT} cause spin-flips and create an effective in-plane magnetic field $[\Delta_{LT} (k_x^2 - k_y^2),2 \Delta_{LT} k_x k_y,0]$. Ref.~\cite{kavokin, physsolidstate1999} extends the result to polaritons taking account the TE-TM splitting of the microcavity. The pseudospin in~\ref{SEqH0} depends on both exciton spin state and its dipole moment orientation, so the spin precession couldn't be observed in empty cavity.

The stationary potential $V_{\text{st}} (\mathbf{r})$ describes the geometry of the structure. It should be emphasized that in general case the potential is complex:
$V_{\text{st}} (\mathbf{r}) = V_{\text{Re}} (\mathbf{r}) + \mathrm{i} V_{\text{Im}} (\mathbf{r})$.
The real part is taken in the azimuthally symmetric form as
\begin{equation}
\label{SEqExtPotents}
V_{\text{Re}} (\mathbf{r}) = V_0 \left\{ \left[ e^{-\frac{(r - r_1)^2}{w^2}} + e^{-\frac{(r - r_2)^2}{w^2}} \right] \Theta (r) +[1- \Theta(r)] \right\},
\end{equation}
where $\Theta(r) = [1-\theta (r-r_2)]\theta(r-r_1)$; $\theta (r)$ is the Heaviside step function,
$r_1$ and $r_2$ are the inner and outer radii of the ring.
The height of the potential $V_0$ and the parameter $w$ are the fitting parameters.
The imaginary part of the stationary potential $V_{\text{Im}} (\mathbf{r})$ is responsible for the leakage of polaritons from the ring due to damping induced by etching~\cite{PhysRevLett126075302,ResInOpt4100105}.
It can be obtained from~\eqref{SEqExtPotents} by replacing $V_0 \rightarrow -V_0 '$ and $r_{1,2} \rightarrow r_{1,2} '$.

The operator
\begin{equation}
\label{SEqEffPotential}
\hat{V} _{\text{eff}} = \hat{V} _{\text{PP}} + \hat{V} _{\text{RP}}
 + \hat{V} _{\text{PD}}
 \end{equation}
describes the effective potential induced by interactions.
The first term in~\eqref{SEqEffPotential} given as
\begin{equation}
\label{SEqEffPotentsPP}
\hat{V} _{\text{PP}} = \frac{1}{2} \alpha  [ n (t, \mathbf{r})  \hat {\sigma}_ 0 + 2 S_z (t, \mathbf{r})  \hat {\sigma}_ z ]
\end{equation}
is responsible for polariton-polariton interactions.
$\alpha$ is the constant of interaction of polaritons with parallel spins.
Interaction of polaritons with opposite spins is weak, and we neglect it.
$n(t,\mathbf{r}) = \Psi ^{\dagger} \Psi = |\Psi _+ (t,\mathbf{r})|^2 + |\Psi _- (t,\mathbf{r})|^2 $ is the total density of polaritons.
$S_z (t,\mathbf{r}) = \Psi ^{\dagger} \hat{S}_z \Psi =\frac{1}{2} [  |\Psi _+ (t,\mathbf{r})|^2 - |\Psi _- (t,\mathbf{r})|^2 ]$  describes the imbalance in the densities of the spin up and spin down states.

The second term in~\eqref{SEqEffPotentsPP} is responsible for interactions of polaritons with bright excitons in the reservoir.
$\hat{V} _{\text{PR}}$ can be obtained from~\eqref{SEqEffPotentsPP} by replacing $n\rightarrow n_{\text{R}}$, $S_z \rightarrow S_{\text{R},z}$ and $\alpha \rightarrow \alpha _{\mathrm{R}}$, where $\alpha _{\mathrm{R}}$ is the polariton-exciton interaction constant:
\begin{equation}
\label{SEqEffPotentsPR}
\hat{V} _{\text{PR}} = \frac{1}{2} \alpha  _{\mathrm{R}} [ n _{\mathrm{R}} (t, \mathbf{r})  \hat {\sigma}_ 0 + 2 S_{\mathrm{R}, z} (t, \mathbf{r})  \hat {\sigma}_ z ] + \hat{\mu} _{\mathrm{R}}.
\end{equation}
In~\eqref{SEqEffPotentsPR}, the operator $\hat{\mu} _{\mathrm{R}}$ is responsible for the effect of flattening the potential profile due to the spreading of the exciton reservoir around the ring, see~\cite{PhysRevB100245304}.
It is treated as an effective chemical potential of the excitons within the ring, $\mu _{\mathrm{R}}$, that increases as density increases (with the increasing pump power).
It is convenient to model it via renormalization of the real component of the stationary potential as following:
\begin{equation}
\label{SEqExtPotentsRenorm}
\hat{V} _{\text{Re}} + \hat{\mu} _{\mathrm{R}} = \text{Max}[V_{\mathrm{Re}} (\mathbf{r}), \mu _{\mathrm{R}}] \cdot \hat {\sigma}_0
\end{equation}

The exciton reservoir is excited by the cw non-resonant optical pump described by the spinor $|P\rangle = P(\mathbf{r}) |p\rangle$, where $|p\rangle = (p_+,p_-)^{\mathrm{T}}$ describes polarization of the pump.
\begin{equation}
\label{SEqPump}
P(\mathbf{r}) \propto \exp \left\{ \left[ (x - x_{\mathrm{p}})^2 + (y - y_{\mathrm{p}})^2 \right] / w_{\mathrm{p}}^2 \right\}
\end{equation}
is responsible for the spatial distribution of the pump taken in the Gaussian form of width $w_{\mathrm{p}}$ shifted by~$(x_{\mathrm{p}} , y_{\mathrm{p}})$.
Both bright and dark excitons emerge under the optical pump.
We take into account the interaction of the condensate with the reservoir of dark excitons as follows:
\begin{equation}
\label{SEqEffPotentsPD}
\hat{V} _{\text{PD}} = \frac{1}{2} \eta P(\mathbf{r}) \hat{\sigma}_0.
\end{equation}
In~\eqref{SEqEffPotentsPD} we assume that the dark reservoir interacts equally with both the spin up and spin down polariton states.
$\eta$~is the fitting parameter which defines the blueshift coming from the dark reservoir.

The term with parentheses in~\eqref{SEqGGPE} is responsible for the balance of pump and decay.
Polaritons in the condensate decay with the rate~$\gamma$.
The operator $\hat{R}_{\text{in}}$ is responsible for the inflow of particles from the reservoir to the condensate.
It is given as follows:
\begin{equation}
\label{SEqRIn}
\hat{R}_{\text{in}} = \frac{1}{2} R [n_{\mathrm{R}}(t,\mathbf{r}) \hat{\sigma} _0 + 2 S_{\mathrm{R},z}(t,\mathbf{r}) \hat{\sigma} _z],
\end{equation}
where $R$ is the stimulated scattering rate.
In~\eqref{SEqResRateEq}, $\gamma _{\mathrm{R}}$ is the reservoir decay rate.
The operator $\hat{R}_{\text{out}}$ is responsible for the outflow of particles from the reservoir to the condensate.
It can be obtained from~\eqref{SEqRIn} by replacing $n \rightarrow n_{\mathrm{R}}$ and $S _{\mathrm{R},z} \rightarrow S_{z}$.

We use the following values of the parameters for simulations.
The effective mass is $m^* = 6.14 \times 10^{-5} m_{\mathrm{e}}$, where $m_{\mathrm{e}}$ is the free electron mass,
the polariton and reservoir decay rates are $\gamma = 1/150 \, \text{ps}^{-1}$ and $\gamma _{\mathrm{R}} = 1/400 \, \text{ps}^{-1}$,
the stimulated scattering rate is $\hbar R = 0.08 \, \text{meV} \, \mu \text{m}^2 $.
The interaction constants are $\alpha = 3 \, \mu \text{eV} \, \mu \text{m}^2$ and $\alpha _{\mathrm{R}} = 10 \, \mu \text{eV} \, \mu \text{m}^2$.
The TE-TM splitting constant is~$\hbar \Delta = 62 \, \mu \text{eV} \, \mu \text{m}^2$.
The parameters of the stationary potential: $V_0 = 3 \, \mu \text{eV}$, $w=3 \, \mu\text{m}$, $r_1 = 40.7\, \mu\text{m}$, $r_2 = 59.3\, \mu\text{m}$,
$V_0' = 1 \, \mu \text{eV}$, $r_1 ' = 40\, \mu\text{m}$, $r_2 ' = 60\, \mu\text{m}$.
For the dark reservoir~$\eta \gamma _{\mathrm{R}} / \alpha _{\mathrm{R}} = 3$.
The polarization of the pump is taken linear,~$|p\rangle = (0.5,0.5)^{\mathrm{T}}$.

\begin{figure}[t]
    \centering
    \includegraphics[scale = 0.45]{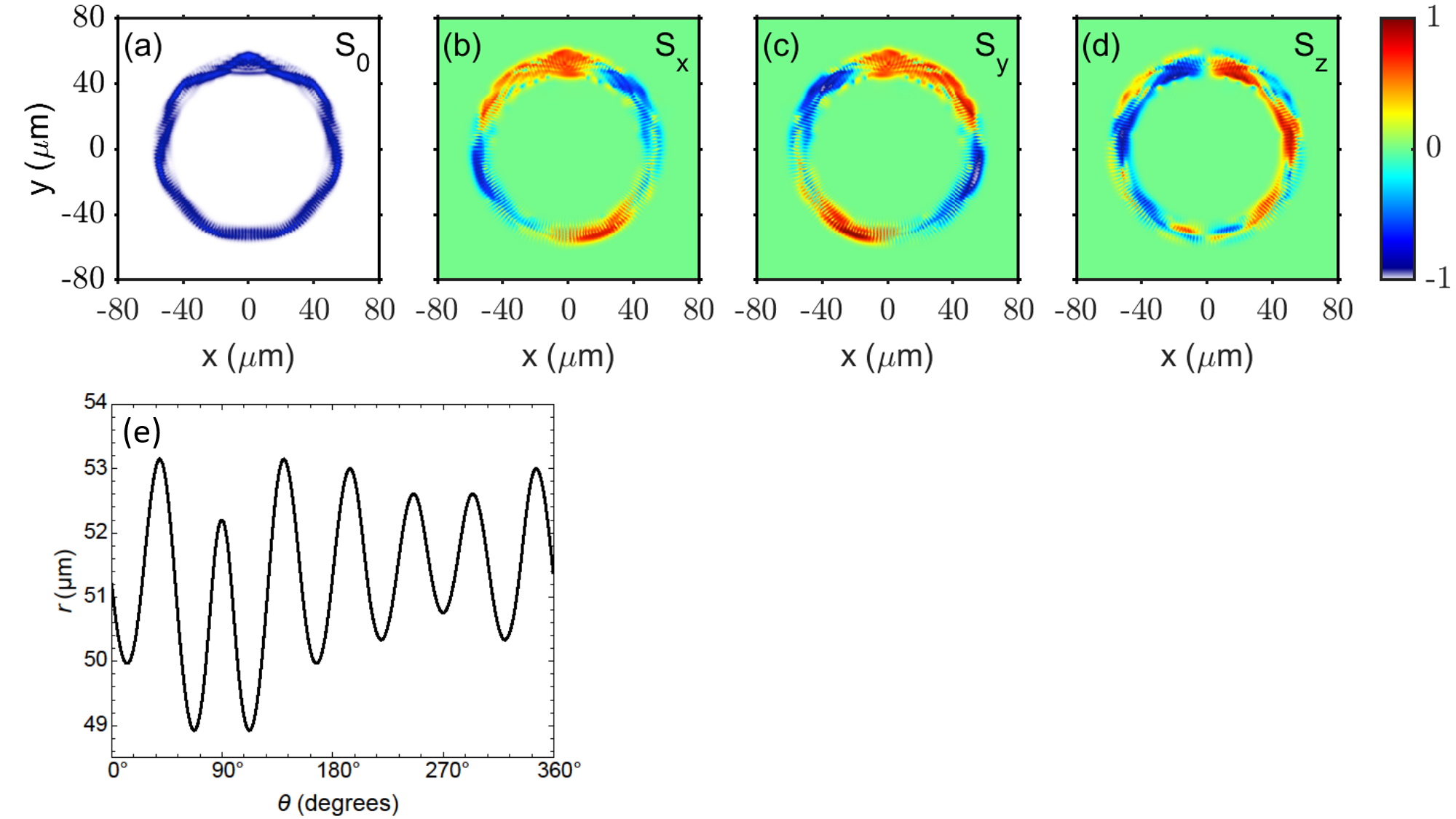}
    \includegraphics[scale = 0.45]{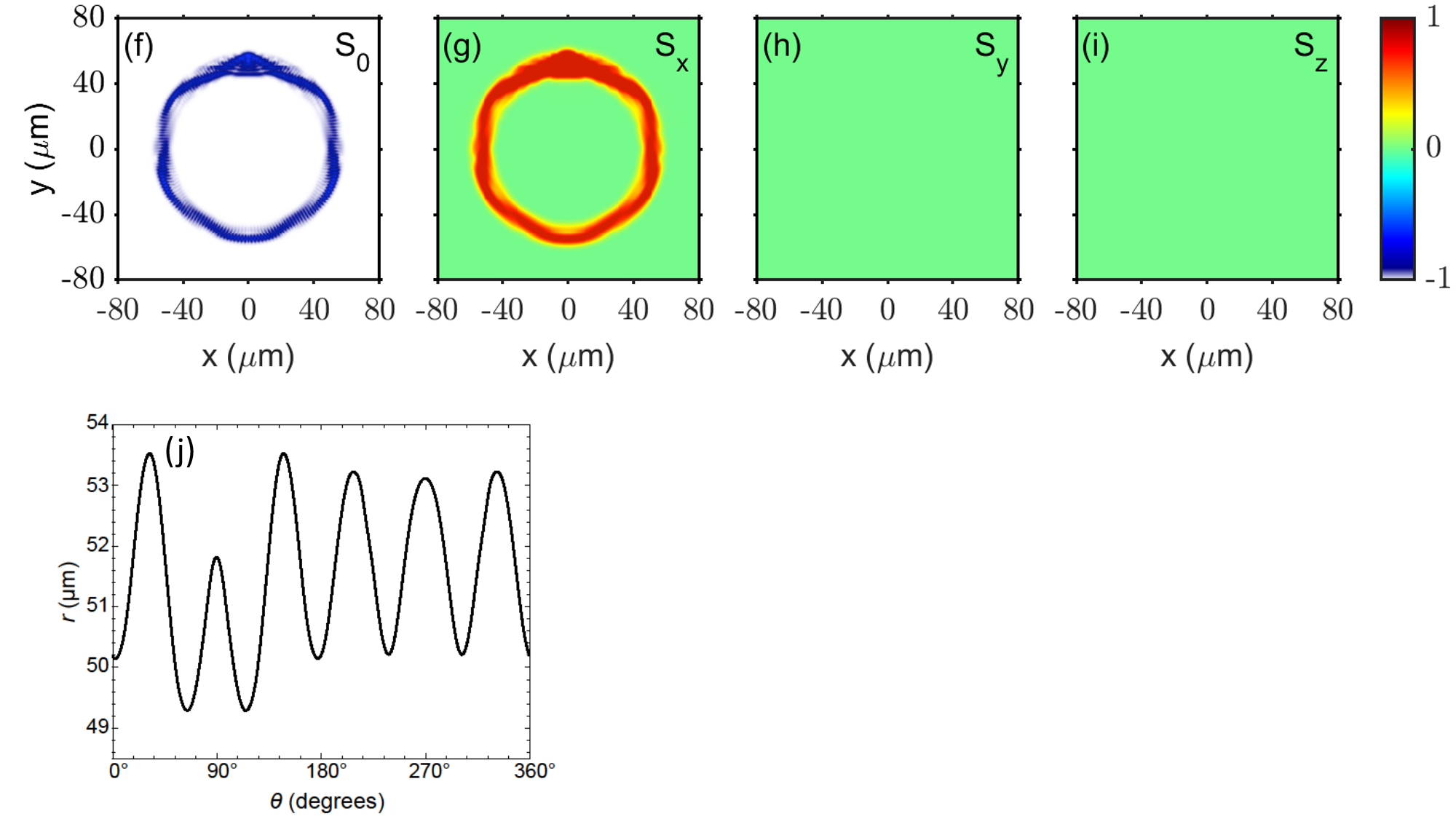}
    \caption{Simulation results correspond to Fig. 3(a) in the main text. (a) conventional intensity distribution. (b)--(d) the three Stokes polarization components. (e) radial center-of-mass coordinate on the polar angle. (f)--(j) simulation results without spin-orbit term, other parameters are the same as (a)--(e).}
    \label{fig:splNOspl1}
\end{figure}

In order to see how SOI affects the intensity distribution and polarization, we compared simulations with and without SOI. Figure \ref{fig:splNOspl1} shows the simulation results corresponding to Fig.~3(a). Panels~\ref{fig:splNOspl1}(a)--(e) are results with SOI, and \ref{fig:splNOspl1}(f)--(j) are results without SOI.
Comparing the panels~\ref{fig:splNOspl1}(e) and~\ref{fig:splNOspl1}(j), which show the variation of the radial center-of-mass coordinate with the polar angle, one can see that SOI contributes to its period of oscillations: seven full periods on a ring in the presence of SOI vs. six periods without SOI.

\begin{figure}[t]
    \centering
    \includegraphics[scale = 0.45]{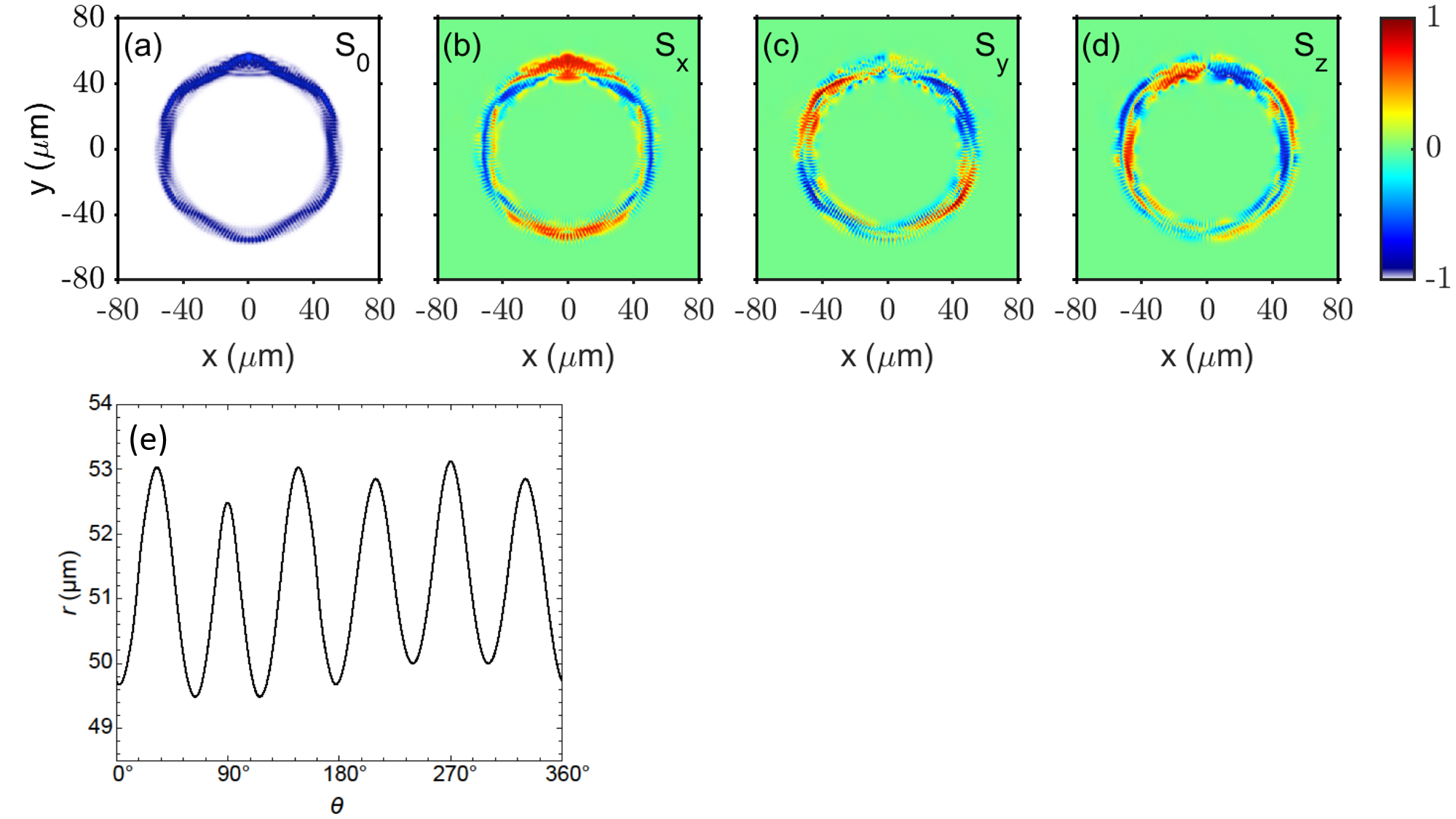}
    \includegraphics[scale = 0.45]{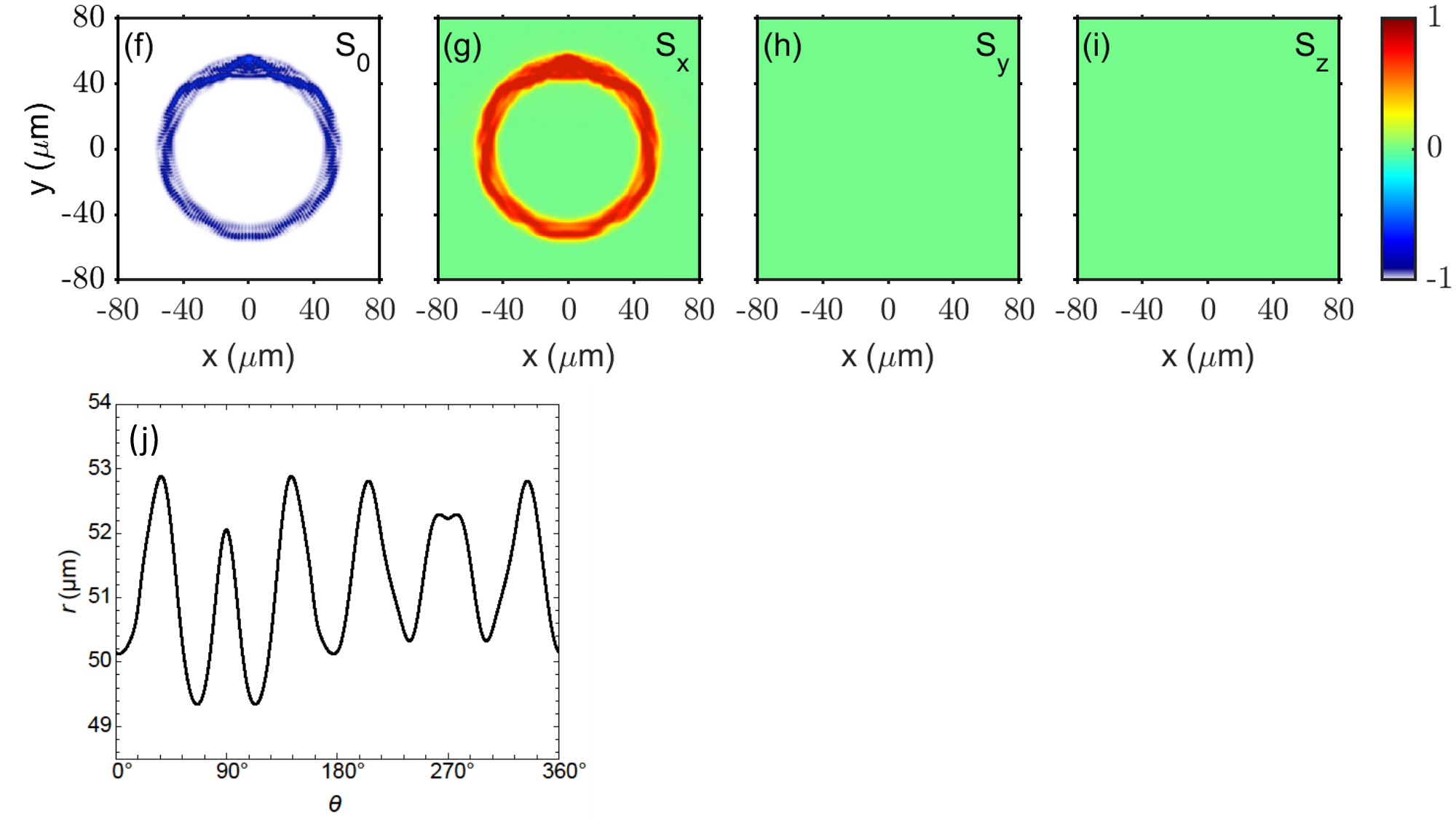}
    \caption{Simulation results correspond to Fig. 3(b) in the main text. (a) conventional intensity distribution. (b)--(d) the three Stokes polarization components. (e) radial center-of-mass coordinate on the polar angle. (f)--(j) simulation results without spin-orbit term, other parameters are the same as (a)--(e).}
    \label{fig:splNOspl2}
\end{figure}

Figure~\ref{fig:splNOspl2} shows a similar comparison corresponding to Fig.~3(b).
Here the decrease in the period of oscillations due to SOI is less remarkable: instead of an additional full period on a ring we have ``doubling'' of a maximum around $270 ^{\circ}$.
Such difference can be justified by different magnitude of SOI due to different wave vectors of polaritons emerging at different pump power, together with periodic boundary conditions for the polariton wave function in a ring geometry, that allow it to be single-determined.


